\documentclass[apj]{emulateapj}
\begin{document}

\title{Lifetime and spectral evolution of a magma ocean with a steam atmosphere: its detectability by future direct imaging}
\author{Keiko Hamano, Hajime Kawahara, Yutaka Abe}
\affil{Department of Earth and Planetary Science, The University of Tokyo,7-3-1 Hongo, Bunkyo-ku, Tokyo 113-0033, Japan}
\email{Correspondence: keiko@eps.s.u-tokyo.ac.jp}

\author{Masanori Onishi}
\affil{Department of Earth and Planetary Sciences, Kobe University,1-1 Rokkodai-cho, Nada, Kobe 657-8501, Japan}

\and
\author{George L. Hashimoto}
\affil{Department of Earth Sciences, Okayama University, 3-1-1 Tsushima-Naka, Kita, Okayama, 700-8530, Japan}

\begin{abstract}
We present the thermal evolution and emergent spectra of solidifying terrestrial planets along with the formation of steam atmospheres. The lifetime of a magma ocean and its spectra through a steam atmosphere depends on the orbital distance of the planet from the host star. For a type-I planet, which is formed beyond a certain critical distance from the host star, the thermal emission declines on a timescale shorter than approximately $10^6$ years. Therefore, young stars should be targets when searching for molten planets in this orbital region. In contrast, a type-II planet, which is formed inside the critical distance, will emit significant thermal radiation from near-infrared atmospheric windows during the entire lifetime of the magma ocean. The Ks and L bands will be favorable for future direct imaging because the planet-to-star contrasts of these bands are higher than approximately $10^{-7}$-$10^{-8}$. Our model predicts that, in the type-II orbital region, molten planets would be present over the main sequence of the G-type host star if the initial bulk content of water exceeds approximately 1 wt\%. In visible atmospheric windows, the contrasts of the thermal emission drop below $10^{-10}$ in less than $10^{5}$ years, whereas those of the reflected light remain $10^{-10}$ for both types of planets. Since the contrast level is comparable to those of reflected light from Earth-sized planets in the habitable zone, the visible reflected light from molten planets also provides a promising target for direct imaging with future ground- and space-based telescopes.
\end{abstract}

\keywords{infrared: planetary systems, planets and satellites: terrestrial planets, radiative transfer}

\section{Introduction}
Theoretical studies predict that the final stage of terrestrial planet formation involves a series of giant impacts between protoplanets. This stage is triggered by the dispersal of nebula gas, and probably lasts approximately 10 to 100 Myr \citep[e.g.][]{Agnor99, Chambers01}. In the aftermath of the last impact, the planets probably begin in a molten state \citep[e.g.][]{Canup04}. Detecting molten terrestrial planets in extrasolar systems is of great significance in testing the widely accepted view of their hot origins. 

The rocky planets can acquire water throughout formation by various processes, such as the local accretion of hydrated and icy planetesimals \citep{Muralidharan08, King10, MA10}, chemical reactions between nebula gas and silicate melts \citep{Sasaki90, IG06}, and delivery via ice-bearing bodies from the outer orbital region \citep{Morbidelli00, Raymond07, Walsh11}. A substantial amount of water would be delivered during planetary formation. Water acquired through the lattermost process alone would exceed the current Earth's ocean mass by more than tens times. Since giant impacts do not blow off the entire water inventory \citep{GA05}, early molten planets are likely to be covered by a steam-rich atmosphere.

Emergent spectra of hot molten planets have been investigated for several atmospheric compositions, including water-rich atmospheres. \citet{M-R09} first presented thermal emission spectra for the hot surface conditions, considering various possible atmospheric compositions; solar and enhanced metallicity atmospheres, $\mathrm{H_2O}$-$\mathrm{CO_2}$ atmospheres, and Venus' composition atmospheres. Since the solidus temperature of peridotite rocks is approximately 1,400 K at the planetary surface \citep[e.g.][]{Takahashi86}, the molten surface emits strong radiation not only in the infrared (IR), but also in the visible and near-IR. They showed that, although the steam-rich atmosphere obscures the hot surface, intense thermal emission from the surface would leak through near-IR atmospheric windows, which are relatively transparent. \citet{Lupu14} further explored emergent high-resolution spectra for atmospheric compositions in equilibrium with magmas with compositions of Earth's continental crust and the bulk silicate Earth, taking into account atmospheric chemistry and structure self-consistently. They found out that the atmospheres are dominated by $\mathrm{H_2O}$ and $\mathrm{CO_2}$, considering quenching effects. Additional opacity sources, such as $\mathrm{CH_4}$ and $\mathrm{NH_3}$, could reduce the brightness temperature by a factor of two to three, but the emergent spectra still reveal the features similar to those without these additional opacity sources. \citet{Lupu14} concluded that the most favorable wavelength range for direct imaging with future telescopes will be 1 to 4 $\mathrm{\mu m}$. 

These previous studies revealed that the mass of the steam atmosphere is a determining factor for spectral detectability of hot molten planets. However, the mass of the steam atmosphere does not remain constant during the solidification process. The atmosphere would grow by degassing from the planetary interior or would escape into space \citep{E-T11, Lebrun13,Hamano13}. Atmospheric evolution would significantly affect the emergent spectra during solidification of the magma ocean. 

In addition, atmospheric evolution is also crucial to the thermal evolution of magma oceans \citep{MA86a, MA86b, Zahnle88,E-T08, E-T11, Lebrun13, Hamano13}. Since the strong greenhouse effect of water vapor prevents heat from escaping, a magma ocean is sustained longer under a more massive atmosphere. Although a growing atmosphere keeps the planet in a molten state for a longer period of time, its spectral signatures become fainter as a result of increasing surface obscuration. Conversely, if the atmosphere escapes with time, the planet becomes brighter, but the magma ocean quickly solidifies over a shorter timescale. Thus, a trade-off relationship is expected between the spectral intensity and occurrence rate of the magma-covered planets, as suggested by \citet{M-R09}.

Variations in water inventory would also affect the planetary spectra and their time evolution. The water in the current Earth's ocean amounts to approximately 0.023 wt \% in bulk content, and its total, including the water stored in the crust and mantle, is probably at most 0.23 wt \% \citep[e.g.][]{Ohtani05, Hirschmann06}. Extrasolar terrestrial planets could exhibit large variations in water amount just after formation as a result of a combination of water supply and loss processes. Recent calculations have indicated that the lifetime of magma oceans depends on the initial water inventory and the orbital distance from the host star \citep{Hamano13}. Therefore, the occurrence rate of hot molten planets as a function of orbital distance would be related to the expected variety in initial water endowment on terrestrial planets.

As described above, spectral calculations of hot molten planets have been presented for various atmospheric compositions by several authors \citep{M-R09, Lupu14}. They have concluded that the detectability in terms of spectral intensities depends primarily on the mass of the atmosphere. On the other hand, recent evolutionary calculations have shown that the mass of the steam atmosphere overlying a magma ocean greatly varies with time by degassing and escape processes \citep{Hamano13}. They also indicated that the evolutionary history of the steam atmosphere and its cooling time strongly depends on the orbital distance from the star. Although they suggested that the occurrence rate of molten planets depends on the planetary orbit as well, their evolutionary model is not able to predict the variations of thermal emission spectra along with the evolution of the steam atmosphere because of the assumption of a gray gas in their radiative transfer model. 

In the present paper, we incorporate a newly developed non-gray radiative transfer model into the evolutionary model by \citet{Hamano13} to describe the spectral evolution of hot molten planets along with the evolution of steam atmospheres during the solidification of a magma ocean. This enables us to self-consistently examine the brightness variations of solidifying planets and the duration of the molten state. We also investigate the dependence of the brightness of solidifying planets on orbital distance and obtain a consistent relationship between planet-to-star contrast and orbital separation. In this paper, we consider Earth-sized terrestrial planets around a Sun-like star.

In Section \ref{sec:model}, we summarize our proposed model and its parameters. We describe the spectral features of thermal emission and the albedo of molten planets with a steam atmosphere. In Section \ref{sec:evol}, we describe the time evolution of surface temperature, atmospheric pressure, and thermal emission spectra. In Section \ref{sec:detect}, we present diagrams of planet-to-star contrast and orbital distance for near-IR atmospheric windows and discuss future prospects for direct imaging.  We also address the possibility of putting constraints on water inventory originating from the formation stages of terrestrial planets from the predicted occurrence rate. Color variations of the solidifying planets are presented as well. In Section \ref{sec:discuss}, we compare our model spectra with those obtained by previous models and discuss parameter uncertainties. In Section \ref{sec:conclusion}, we summarize our main findings and present our conclusions.

\begin{figure*}[tbp]
\epsscale{1}
\plotone{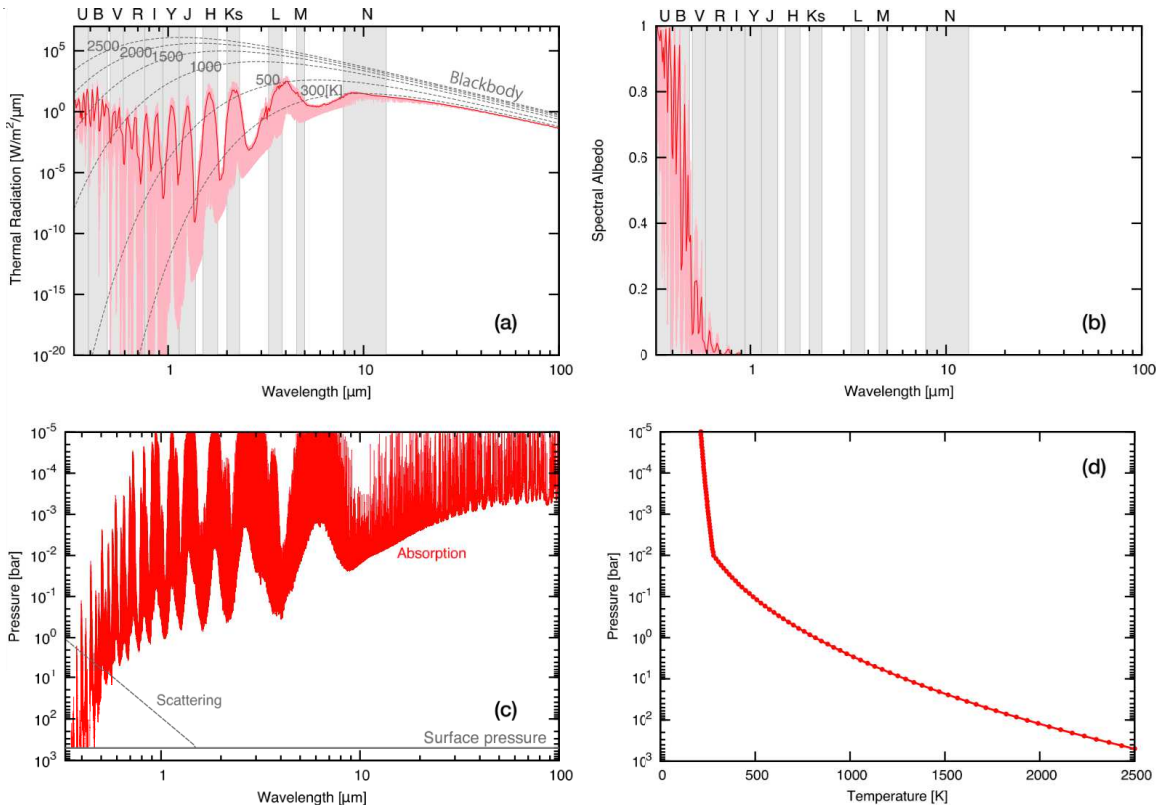}
\caption{(a) Thermal emission spectrum through a steam atmosphere of 50 bar with a surface temperature of 2,500 K. The light-pink line indicates a high-resolution spectrum with a wavenumber resolution of 0.01 $\mathrm{cm^{-1}}$, corresponding to a resolving power $R_\lambda$ of approximately 1,000,000 at 1 $\mathrm{\mu m}$. The red line indicates the average thermal radiation such that $R_\lambda$ becomes 100. Gray bands represent atmospheric windows, which are defined in Table \ref{tab1}. (b) High-resolution (pink) and average (red) spectral albedo for a stellar zenith angle $\theta_{\bigstar} $ of $\cos^{-1}(1/\sqrt{3})$. (c) Pressure level at which the optical depth is unity for absorption (red line) and for scattering (dotted gray line). (d) Atmospheric structure.}
\label{fig:LBLspec}
\end{figure*}

\section{Model description and spectral features}\label{sec:model}
\subsection{Radiative transfer calculations}
The evolutionary model used in the present study is that described in \citet{Hamano13}, except for the calculations of radiative transfer. In the present study, we replace a gray radiative transfer code used in \citet{Hamano13} with a newly developed, non-gray code to calculate thermal emission and reflection spectra from the planet. In the non-gray code, a two-stream formulation is used to calculate the radiative transfer in a plane-parallel atmosphere. We calculate planetary and stellar radiation separately in a wavenumber range from 0 to 30,000 cm$^{-1}$ with a spectral resolution of 0.01 cm$^{-1}$. The albedo of the planetary surface is fixed at 0.2. See the Appendix for details on the radiative transfer model and its validity. 

A steam atmosphere is divided into 100 layers, which are equally spaced on a logarithmic scale of pressure. The pressure at the top of the atmosphere is fixed at 1 Pa in all calculations. The temperature structure is prescribed such that it follows an adiabatic lapse rate in all atmospheric layers, based on results of \citet{AM88}. For atmospheric layers in which the temperature is above the critical temperature of water or the relative humidity is less than 1, the lapse rate is assumed to be a dry adiabat. For the other layers, the lapse rate is given by a pseudo-moist adiabat under the assumption that condensates are removed immediately as they form. For simplicity, the effects of clouds are ignored in the present study. Since we consider a pure steam atmosphere, the pseudo-moist adiabat overlaps the saturation curve of water in the present study. We apply a liquid-vapor saturation curve at temperatures above 273 K and an ice-vapor curve at temperatures below 273 K. The dry adiabat is calculated using the Peng-Robinson equation of state in a manner similar to \citet{AM88}, but the heat capacity of water vapor in the ideal gas state is taken from \citet{WP02} as a function of temperature in order to make the heat capacity applicable to higher-temperature conditions.

We use HITEMP 2010 \citep{hitemp} as spectral database to calculate the absorption coefficients for lines of water vapor. For all calculations, a Voigt line profile is used to consider the combined effect of Doppler and pressure broadening, using a computational code by \citet{Humlicek82}. Although no reliable model for water continuum is available for the high-temperature, high-pressure conditions considered herein, we adopt a MT\_CKD 2.5 \citep{Mlawer12} model for its continuum together with a wavenumber truncation of 25 cm$^{-1}$, which is similar to a procedure used in recent studies on radiative transfer calculations of hot and thick steam atmospheres \citep{Kopparapu13, Goldblatt13}. As the number of line transitions in the HITEMP 2010 exceeds $10^8$, we generate a grid of absorption coefficients of water vapor and calculate the optical depth of each layer by interpolating the absorption coefficients. 

We consider Rayleigh scattering by water vapor with a scattering cross-section per molecule of
\begin{eqnarray}
	\sigma_\mathrm{sc} &=& \frac{32 \pi^3}{3}\frac{1}{N_\mathrm{A}^2} \left( \frac{m_0-1}{n_0} \right)^2 W_\mathrm{n}^4,
	\label{eq:RTsc}
\end{eqnarray}
assuming that the polarizability volume of water vapor does not depend on temperature or pressure. In the above equation, $W_\mathrm{n}$ is the wavenumber, $N_\mathrm{A}$ is the Avogadro constant, and $n_0$ and $m_0$ are the molar volume and the refractive index in the reference state, respectively. We adopt a refractive index of 1.000254 from \citet{Allen73} for D lines of Na at the standard temperature and pressure (0$ ^\circ\mathrm{C}$, 1 atm).
\begin{center}
\begin{deluxetable}{ccc}
	\tablecaption{The band parameters used in the present study.}
	\tablehead{
		\colhead{Band} &\colhead{$\lambda_\mathrm{c}\ \mathrm{[\mu m]}$} &\colhead{$\Delta \lambda\ \mathrm{[\mu m]}$} }
	\tablewidth{200pt}
	\startdata
	U & 0.365 & 0.068 \\
	B & 0.44 & 0.098 \\
	V & 0.55 & 0.089 \\
	R & 0.70 & 0.22 \\
	I & 0.85 & 0.18 \\
	Y & 1.04 & 0.20 \\
	J & 1.26 & 0.24 \\
	H & 1.65 & 0.29 \\
	Ks & 2.16 & 0.32 \\
	L & 3.55 & 0.57 \\
	M & 4.77 & 0.45 \\
	N & 10.47 & 5.19 
	\enddata
	\tablerefs{The band parameters for I, Y, and J are defined by authors, and the other band parameters are taken from \citet{TO10}.}
	\label{tab1}
\end{deluxetable}
\end{center}
	
\subsection{Thermal emission}
Figure \ref{fig:LBLspec}(a) shows the thermal emission spectrum from 0.33 to 100 $\mathrm{\mu m}$ with a surface pressure of 50 bar and a surface temperature of 2,500 K. At this surface temperature, the peak wavelengths of the Planck function $B_\lambda$ and $\lambda B_\lambda$ are approximately 1.2 and 2 $\mathrm{\mu m}$, respectively. The planetary surface emits strong radiation in the visible and near-IR. Although the hot planetary surface emits significant radiation, the outgoing thermal emission from the top of the atmosphere is greatly attenuated by absorption and scattering by the steam atmosphere. The stronger thermal radiation passes through the wavelength regions corresponding to the atmospheric windows, at which the absorption of water vapor is relatively weak.

Figure \ref{fig:LBLspec}(c) shows the atmospheric pressures at which the optical depth is unity for absorption and scattering, respectively. In the far-IR and absorption bands of water vapor, the upper layers, in which the pressure is typically less than $10^{-3}$ bar, are responsible for the thermal emission spectrum, whereas the atmospheric windows reveal the deeper and hotter layers. At wavelengths longward of approximately 0.6 $\mu$m, the pressure level at which the optical depth of absorption is unity is much lower than that of scattering. In the near-IR wavelength region, therefore, the brightness temperature provides a good estimate of the atmospheric temperature at an optical depth of unity. 

At the visible wavelength, the scattering process has a dominant role in attenuating the thermal emission. At shorter wavelengths, the deepest layer could contribute to the emergent spectrum because of the absence of strong absorption lines. The actual probed depths, however, depend on the atmospheric amount and continuum absorption of water vapor, the physics of which remain poorly understood. If the atmosphere is thin enough to allow the thermal radiation emitted from the surface to pass through the atmosphere, a Rayleigh scattering slope appears in the planetary spectrum at sufficiently optically thin wavelengths. Measuring the scattering slope can, in principle, provide information about the atmospheric column mass, the surface temperature and the surface albedo.

\subsection{Spectral albedo}
Figure \ref{fig:LBLspec}(b) shows the spectral albedo for a stellar zenith angle $\theta_{\bigstar}$ of $\cos^{-1} \left( 1/\sqrt{3}\right)$. The spectral albedo is defined as the ratio of the upward diffusive radiation to the downward direct radiation at the top of the atmosphere. The bond albedo of the planet is obtained as a weighted mean of the spectral albedo by the Planck function with an effective stellar temperature.

In addition to the thermal radiation, the spectral albedo is higher within the atmospheric windows. Moreover, the spectral albedo varies greatly at the visible wavelengths, from approximately 1 at 0.4 $\mathrm{\mu m}$ to less than 0.1 beyond 0.9 $\mathrm{\mu m}$. Therefore, in the near-IR, the reflected stellar light is expected to have a smaller contribution to the emergent spectrum. Although the spectral and resulting Bond albedos increase with stellar zenith angle, the difference is less than 0.2 throughout the visible wavelengths between $\theta_{\bigstar}$ of 0 and 80 degrees. In the present study, we use the spectral and Bond albedos with $\theta_{\bigstar}$ of  $\cos^{-1} \left( 1/\sqrt{3}\right)$ degrees as a representative value. Spectra composed of both reflections and emissions are presented later herein under conditions that are consistent with the atmospheric evolution in the course of solidification of the magma ocean.

\subsection{Evolutionary model}
Energy and water budgets during planetary solidification are treated in the same manner as \citet{Hamano13}. Here, we briefly summarize our proposed model, focusing especially on the major changes made in the present study. For additional details, see \citet{Hamano13}. We assume an adiabatic temperature profile in the planetary interior to obtain heat capacity of the whole silicate portion. The heat flux from the magma ocean is calculated as the difference between the outgoing planetary radiation and the net incoming stellar radiation. In order to reduce the computing time for the evolution calculations, the planetary radiation and albedo are precalculated on 882 surface-temperature-pressure grid points, from 1,350 to 3,000 K and from $5\times 10^{-4}$ to 5,000 bar, respectively. In order to follow the evolution of planet-to-star contrast, the thermal radiation from each band is also tabulated after averaging the high-resolution spectra with a resolving power of 100, which is a typical value for future direct imaging. We consider a Sun-like host star, the bolometric luminosity and XUV radiation of which evolve in a manner similar to solar analogs. We adopt a solar standard model by \citet{Gough81} for luminosity enhancement with stellar age and linearly extrapolate it after a stellar age of 4.567 Gyr. The stellar spectrum is assumed to be a blackbody spectrum with an effective stellar temperature of 5,800 K.

\begin{figure*}[tbp]
\epsscale{1}
\plotone{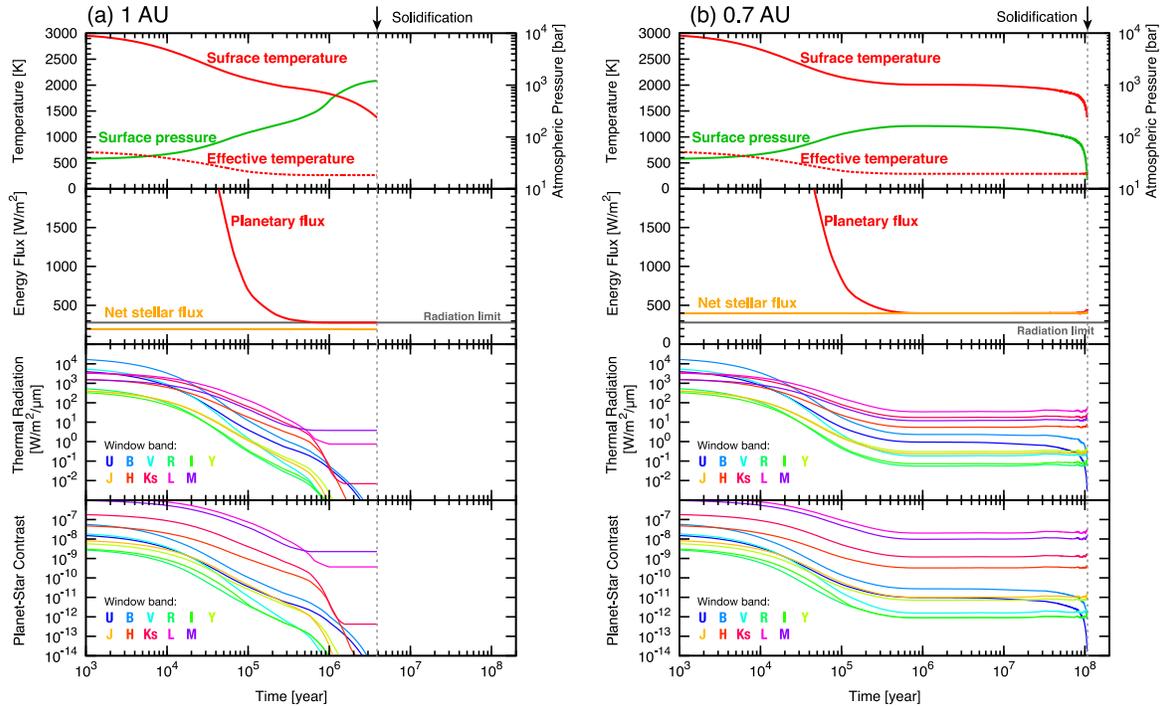}
\caption{Evolution of a planet at (a) 1 AU (type I) and (b) 0.7 AU (type II) with an initial water mass of 5 $M_\mathrm{EO}$. The third and bottom panels, respectively, show the thermal radiation and the maximum value of the planet-to-star contrast within each atmospheric window. The dotted vertical line in each figure indicates the time at which the magma ocean has solidified, which is approximately 3.9 Myr for the type-I planet and 100 Myr for the type-II planet.}
\label{typeI&II_evol}
\end{figure*}

Water is assumed to be partitioned between the atmosphere and the magma ocean according to its solubility into basaltic melts. The incorporation of water into cumulates is also considered according to the partition coefficients of water between melts and solids and by assuming 1\% of trapped interstitial melts. The mass loss rate of water is calculated from an energy-limited escape flux of hydrogen $\Gamma_\mathrm{H}$ \citep{Watson81} with a heating efficiency $\eta$ of 0.1, as follows:
\begin{eqnarray}
	\Gamma_\mathrm{H} &=& 4\pi R_\mathrm{pl}^2 \frac{R_\mathrm{pl}}{G M_\mathrm{pl}} \frac{\eta S^\mathrm{XUV}_\bigstar}{4} \left( \frac{a}{\mathrm{1\ AU}} \right)^{-2}, \label{eq:loss_rate}
\end{eqnarray}
where $R_\mathrm{pl}$ and $M_\mathrm{pl}$ are the planetary radius and mass, respectively, $a$ is the orbital distance from the host star in AU, and $G$ the gravitational constant. Here, $S^\mathrm{XUV}_\bigstar$ is the XUV flux from the host star at 1 AU. The stellar XUV flux decreases rapidly with time, whereas the ratio of the X-ray luminosity to the bolometric luminosity remains nearly constant at $\sim 10^{-3}$ for young active stars \citep{VW87,Stauffer94, Pizzolato03}. Based on a power-law relation derived from observational data by \citet{Ribas05} and \citet{Ribas09}, we adopt the following expression for the evolution of $S^\mathrm{XUV}_\bigstar$ as a function of stellar age $\tau$ in Gyr:
\begin{eqnarray}
   S^\mathrm{XUV}_\bigstar = \left\{ \begin{array}{ll}
   	\displaystyle{S^\mathrm{XUV}_\odot \left( \frac{\tau_\mathrm{sat}}{\tau_\odot} \right)^{{-1.23}}}  & \mathrm{for\ } \tau < \tau_\mathrm{sat} \\
	\displaystyle{S^\mathrm{XUV}_\odot \left( \frac{\tau}{\tau_\odot} \right)^{-1.23}}  & \mathrm{for\ } \tau > \tau_\mathrm{sat}
  \end{array} \right.
  \label{eq:XUV}
\end{eqnarray}
and
\begin{eqnarray}
	\tau_\mathrm{sat} &\equiv& 1.66\times10^{20} L_\mathrm{bol}^{-0.64}, 
\end{eqnarray}
where $L_\mathrm{bol}$ is the bolometric luminosity of the host star in $\mathrm{erg\ s^{-1}}$, and $\tau_\odot$ and $S^\mathrm{XUV}_\odot$, which are the age and XUV flux at 1 AU of the current Sun, are fixed to 4.567 Gyr and $4.59\times 10^{-3}$ $\mathrm{J\ s^{-1}\ m^{-2}}$, respectively. The value of $\tau_\mathrm{sat}$ is approximately 0.066 Gyr for the solar standard model by \citet{Gough81}. The mass loss rate of water is nine times as large as the mass loss rate of hydrogen under the assumption that all the dissociated oxygen atoms are consumed to oxidize the abundant surface magma \citep{Gillmann09, Hamano13}.

We consider terrestrial planets with the same mass and bulk composition as Earth. We start the evolution calculations from a very hot state with a surface temperature of 3,000 K and stop the calculations when the surface temperature reaches the solidus temperature at the surface (approximately 1,370 K) or the stellar age reaches 10 Gyr, which is comparable to the main sequence lifetime of the host star. The solidification is assumed to start at a stellar age of 0.05 Gyr ($\tau_0$). We consider the initial total inventory of water ranging from 0.01 to 50 times the current ocean mass of Earth $M_\mathrm{EO}$ of $1.4\times 10^{21}$ kg. This corresponds to a range in bulk water content of from $2.3\times 10^{-4} $ to 1.2 wt \%. Furthermore, we consider the orbital distance from the star of between 0.4 and 1.5 AU. 

\section{Spectrum evolution during solidification}\label{sec:evol}

A hot planet cools and solidifies as it emits large outgoing radiation. Spectral variations during planetary solidification reflect the evolution of surface temperature and atmospheric pressure. At early stages of evolution, the highest fraction of water is expected to dissolve in a deep magma ocean because of the high solubility of water into silicate melts \citep{Zahnle10}. The steam atmosphere therefore accounts for a modest fraction of its total inventory. As the solidification proceeds, the silicate magma becomes enriched in water and degases at a rate that depends on the solidification rate of the magma ocean \citep[e.g.][]{E-T08}. Whether the steam atmosphere grows with time depends on the net balance of the degassing rate of water from the interior and its loss rate by hydrodynamic escape \citep{Hamano13}.
\begin{figure*}[tb]
\epsscale{1}
\plotone{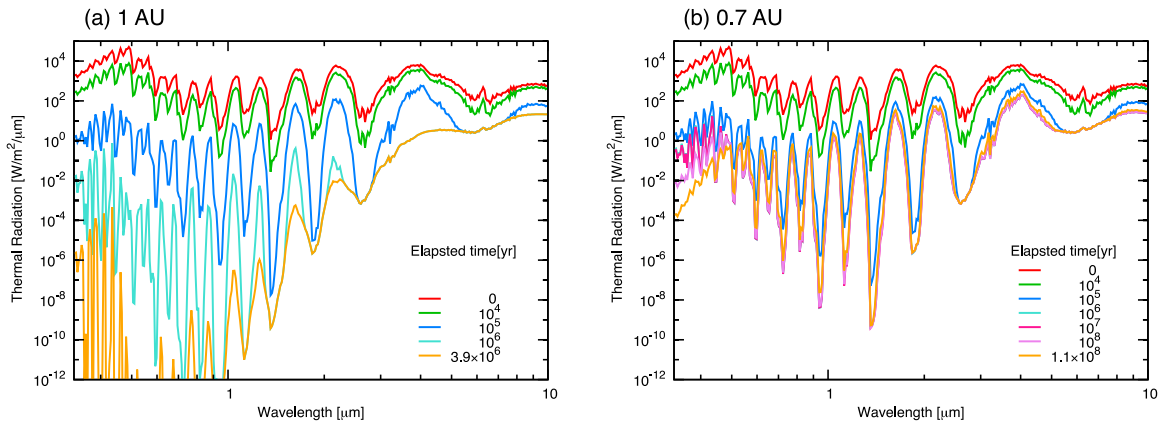}
\caption{Spectral evolution in the visible and near-IR during the course of solidification of the planet at (a) 1 AU (type I) and (b) at 0.7 AU (type II) with an initial water mass of 5 $M_\mathrm{EO}$. The color of the thermal emission spectra represents the elapsed time after solidification starts. The orange line indicates the thermal spectrum at the time at which the planet has solidified.}
\label{typeI&II_spec}
\end{figure*}
Considering a magma ocean covered with a steam atmosphere, \citet{Hamano13} proposed that terrestrial planets have a distinct early evolution that depends on whether the orbital distance from the host star exceeds a critical distance. The critical distance $a_\mathrm{cr}$\footnote{We rescaled the expression for $a_\mathrm{cr}$ in \citet{Hamano13} using the typical values in the present study.} is defined as the distance at which the net stellar heating flux is equal to the radiation limit $F_\mathrm{lim}$ of steam atmospheres:
\begin{eqnarray}
	a_\mathrm{cr} &\approx& 0.83 \left( \frac{F_\mathrm{lim}}{280\, \mathrm{Wm^{-2}}} \right)^{\!\mathalpha{-}1\mathrel{/}2\!} \left( \frac{S^\mathrm{bol}_\mathrm{\bigstar}(\tau_0)}{0.7\, S^\mathrm{bol}_\mathrm{\odot}} \right)^{1/2} \left( \frac{1-\alpha_\mathrm{pl}}{1-0.2} \right)^{1/2}\ \mathrm{AU}
\end{eqnarray}
where $\alpha_\mathrm{pl}$ is the Bond albedo of the planet, and $S^\mathrm{bol}_\mathrm{\bigstar}$ and $S^\mathrm{bol}_\mathrm{\odot}$ are the bolometric fluxes at 1 AU of the host star and the current Sun, respectively.

When the troposphere is saturated by water vapor to a sufficiently deep level, the outgoing planetary flux becomes constant. This constant value is referred to as the radiation limit and corresponds to the lower limit of the planetary radiation for the hot surface conditions considered in the present study. The radiation limit of the steam atmospheres has been extensively investigated, and its value is approximately 280 to 310 $\mathrm{W\, m^{-2}}$ \citep{AM88, Kasting88, Nakajima92,Kopparapu13,Goldblatt13}. Our model yields a radiation limit of 280 $\mathrm{W\, m^{-2}}$. \citet{Hamano13} also classified terrestrial planets into two types according to their orbital distance: type I for planets beyond $a_\mathrm{cr}$ and type II for planets inside $a_\mathrm{cr}$. These two types of planets will solidify over vastly different timescales, and their atmospheres also evolve differently, due to different cooling mechanisms. In the following subsections, we describe the spectral changes over the solidification of type-I and type-II planets with an initial water inventory of 5 $M_\mathrm{EO}$ as a nominal case.

\subsection{Type I:  planets beyond $a_\mathrm{cr}$}\label{ss:typeI}
The type-I planet is characterized by a sufficiently far distance from the host star, where the net incident stellar radiation is smaller than the radiation limit. Since the outgoing planetary radiation never falls below the radiation limit, the type-I planet is self-luminous over its solidification period. The minimum heat flux, which is given by the difference between the radiation limit and the net incident stellar flux, determines the overall solidification timescale, which is approximately 4 Myr in this nominal case (Fig. \ref{typeI&II_evol}(a)). The rapid solidification results in the monotonic growth of the steam atmosphere towards the end of the magma ocean period by keeping the degassing rate of water higher than its loss rate.

Although the surface temperature during the magma ocean period is extremely high, the bolometric luminosity is not so large compared to that from the hot surface. Here, we define an effective emission temperature $T_\mathrm{eff}$, using the planetary radiation $F_\mathrm{pl}$, as follows:
\begin{eqnarray}
	T_\mathrm{eff} &=& \sqrt[4]{F_\mathrm{pl}/\sigma}.
\end{eqnarray}
The effective emission temperature never exceeds 800 K in this nominal case and is typically as low as 300 K over the solidification period. The thick steam atmosphere thus sustains the magma ocean longer, whereas the planet itself looks as faint as a planet with a cool or moderate surface temperature.

The most recognizable diagnostic sign of a hot surface is emission at the visible and near-IR wavelengths. Although the massive steam atmosphere strongly absorbs the thermal emission from hotter and deeper layers, thermal emission can leak through atmospheric windows. The third panel of Fig. \ref{typeI&II_evol}(a) shows the time variations of thermal radiation from atmospheric windows. In the early stages (less than $10^4$ years), the thermal radiation from all bands exceeds $10^{2}\, \mathrm{W\, m^{-2}\, \mu m^{-1}}$. These high fluxes, however, do not last long. At most of the atmospheric windows, the thermal fluxes rapidly decrease with time and fall below $10^{-2}\, \mathrm{W\, m^{-2}\, \mu m^{-1}}$ due to the decrease in the surface temperature and the growth of the steam atmosphere. The thermal radiation of the M, L, and Ks bands initially decreases with time as well, but eventually levels off.

Figure \ref{typeI&II_spec}(a) shows the spectral evolution in the visible and near-IR. The planetary thermal radiation declines with time, and then, at wavelengths longer than approximately 2 $\mathrm{\mu m}$, the thermal radiation becomes constant. This is a result of the same mechanism as the occurrence of the radiation limit of the steam atmospheres and indicates that, for the wavelengths at which water vapor has strong absorption, the emission level becomes as high as that at which the thermal structure is uniquely determined by the saturation curve of water. Since the thermal structure around the emission level remains the same, the thermal radiation from the M, L, and Ks bands becomes constant, as shown in Fig. \ref{typeI&II_evol}(a).

\subsection{Type II: planets inside $a_\mathrm{cr}$}\label{ss:typeII**}
In contrast to the type-I planet, the type-II planet receives a net stellar radiation that is larger than the radiation limit because of its proximity to the host star. The planet therefore would reach a radiative energy balance during the course of solidification. Early in its solidification, the outgoing planetary flux far exceeds the net stellar flux, and the planet is highly self-luminous, as in the case of the type-I planet (Fig. \ref{typeI&II_evol}(b)). In the case of the type-II planet, however, the planetary flux decreases with time and eventually balances the net stellar flux at approximately 0.6 Myr. If no water loss occurs, the radiative balance is perfectly achieved, and the magma ocean does not solidify. Actually, as the steam atmosphere becomes thinner with time due to the hydrodynamic escape, the planet emits slightly larger radiation than the net stellar radiation. The type-II planet thus slowly cools and solidifies, losing its atmosphere.  

The solidification rate of the magma ocean is regulated by the loss rate of water such that a sufficient net loss of the steam atmosphere occurs so that the net outgoing flux becomes positive. The resulting solidification time is approximately 100 Myr in this nominal case, which is much longer than that of the type-I planet. The most striking feature is that the higher band fluxes continue over this long solidification period (third panel in Fig. \ref{typeI&II_evol}(b)). The thermal radiation from the atmospheric windows ceases to decrease at $\sim$ 0.6 Myr, at which radiative energy balance has been roughly achieved. The subsequent spectra do not vary with time at the opaque IR wavelengths or the atmospheric windows, which are relatively transparent (Fig. \ref{typeI&II_spec}(b)). 

\begin{figure}[tbp]
\epsscale{1}
\plotone{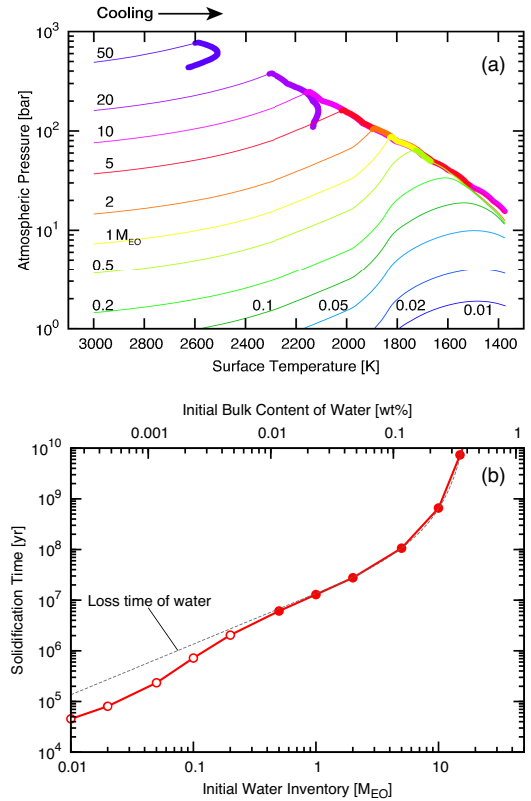}
\caption{(a) $p_\mathrm{s}-T_\mathrm{s}$ evolutionary paths of the planet at 0.7 AU. The number on each curve indicates the initial inventory of water in the current Earth's ocean mass $M_\mathrm{EO}$ of $1.4\times 10^{21}$ kg. The thick parts indicate that the planet is in quasi-energy balance. Here, the `quasi-energy balance state' is defined as the state in which the relative difference between the planetary and net stellar radiation is less than 5\%. The hooks of the $p_\mathrm{s}-T_\mathrm{s}$ curves with an initial water inventory of 20 and 50 $M_\mathrm{EO}$ result from the enhancement of the stellar luminosity. (b) Solidification time of the planet at 0.7 AU as a function of initial water inventory. The filled circles represent the cases in which the quasi-energy balance state has been attained at some point during solidification, and the open circles indicate the cases in which the quasi-energy balance state has not been attained at some point during solidification. The gray dashed line is the time required for complete loss of water.}
\label{typeII_pTpath}
\end{figure}

The long-lived and constant thermal emission in the window regions occurs by attaining quasi-energy balance on the type-II planet. In the quasi-energy balance state, the outgoing planetary radiation roughly balances the incoming net stellar radiation. In order to satisfy this requirement, the amount of the atmosphere must be regulated during the course of solidification. As the surface temperature decreases with time, the amount of the atmosphere must decrease such that the planet maintains a planetary radiation that is nearly equal to, or, strictly speaking, slightly larger than, the net stellar flux (Fig. \ref{typeI&II_evol}(b)). Consequently, the resulting spectra do not change during a quasi-energy balance state, except at the shorter wavelengths, at which the opacity of the atmosphere is thin enough for the outgoing thermal emission to be sensitive to surface temperature (Fig. \ref{typeI&II_spec}(b)). 

The above requirement for planetary radiation generally applies to planets in (quasi-)energy balance, irrespective of the atmospheric amounts and composition. Although additional opacity sources would increase the opacities at some wavelengths and would change the overall spectra, the planet in quasi-energy balance is still expected to emit high thermal radiation from the relatively transparent wavelengths.

Figure \ref{typeII_pTpath}(a) shows the evolutionary paths of surface temperature and pressure (the $p_\mathrm{s}-T_\mathrm{s}$ path) during solidification for various initial inventories of water. The atmospheric pressure increases at early stages due to the high degassing rate. For an initial water inventory larger than 0.2 to 0.5 $M_\mathrm{EO}$, the quasi-energy balance is attained at some particular surface temperature. The $p_\mathrm{s}-T_\mathrm{s}$ paths in the quasi-energy balance states are approximately independent of the initial water endowment, as long as the planets solidify during the main sequence of the host star.

For a larger initial inventory of water, the planet reaches quasi-energy balance at a higher surface temperature and solidification requires a longer period of time. In particular, for the case in which the quasi-energy balance has been achieved during the course of solidification, the lifetime of the magma ocean is well approximated by the time required for total loss of the initial water inventory (Fig. \ref{typeII_pTpath}(b)). If the planet is endowed with water equivalent to more than 16 $M_\mathrm{EO}$,  the molten state would exist for over 10 Gyr, which is comparable to the lifetime of the G-type host star. In this case, the surface temperature gradually increases with time in accordance with the luminosity enhancement of the star.

\begin{figure*}[tbp]
\epsscale{1}
\plotone{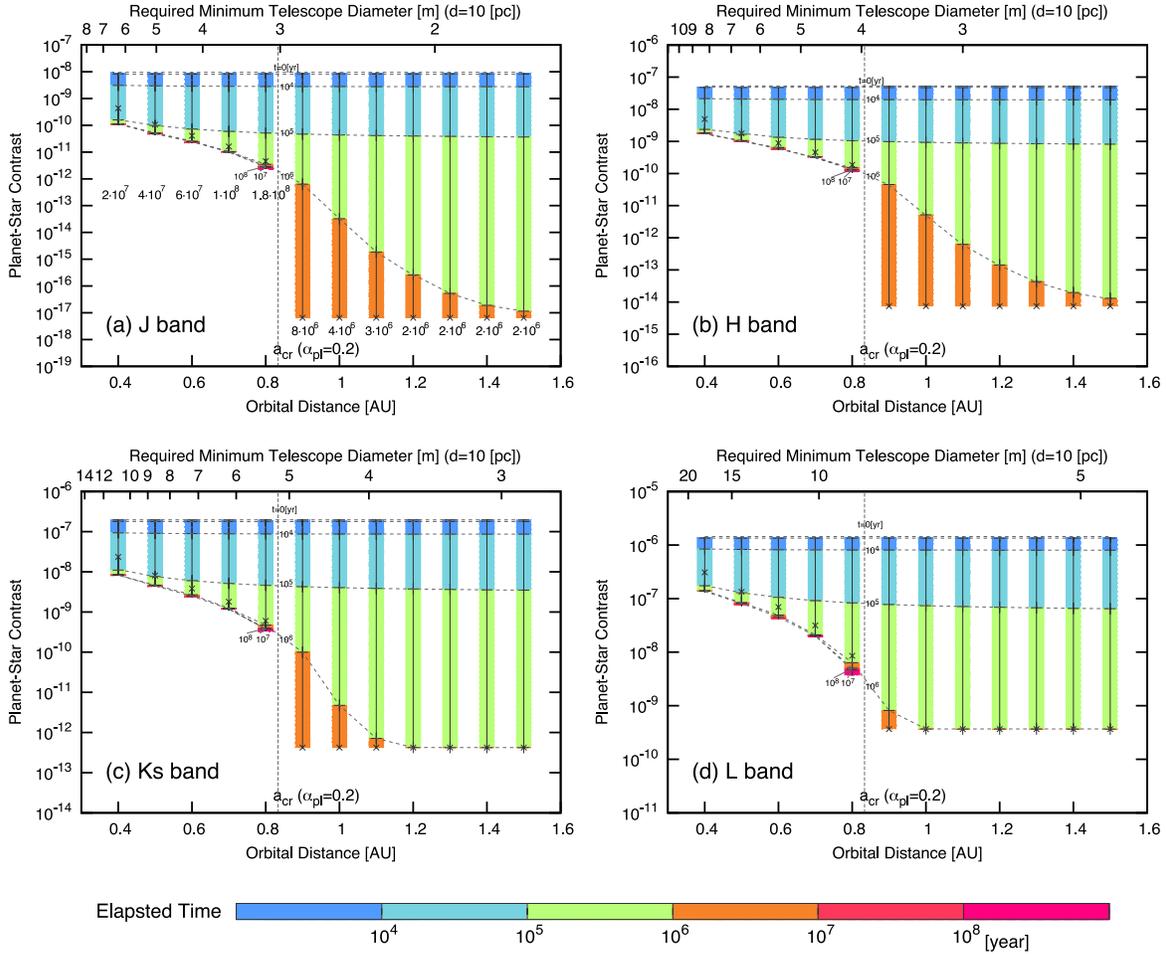}
\caption{Contrast ranges of the thermal emission during the magma-ocean period at each orbital distance with an initial inventory of water of 5 $M_\mathrm{EO}$: (a) J, (b) H, (c) Ks, and (d) L bands. Each color bar indicates a contrast range during solidification, with the tick marks indicating the elapsed time for each order of magnitude. The cross symbols indicate the contrasts at which the magma ocean has completely solidified. The numerical value under each color bar in Fig. \ref{a_contrast}(a) is the overall solidification time. At orbital distance smaller than 0.8 AU, the planet-to-star contrast increases with time at the very last stage of solidification because of the low degassing rate. The increase in the contrast at the very last stage is neglected in this figure due to its short duration relative to the total solidification time. The upper horizontal axis shows the telescope diameter $D_\mathrm{tel}$ at which the angular separation is equal to the diffraction limit, $\theta_\mathrm{dl} \sim 0.01 \times (\lambda_\mathrm{c}/0.5\ \mathrm{\mu m}) \times (D_\mathrm{tel}/10\ \mathrm{m})^{-1}$, for the case in which the distance to the system $d$ is 10 pc.}
\label{a_contrast}
\end{figure*}

\begin{figure*}
\epsscale{1}
\plotone{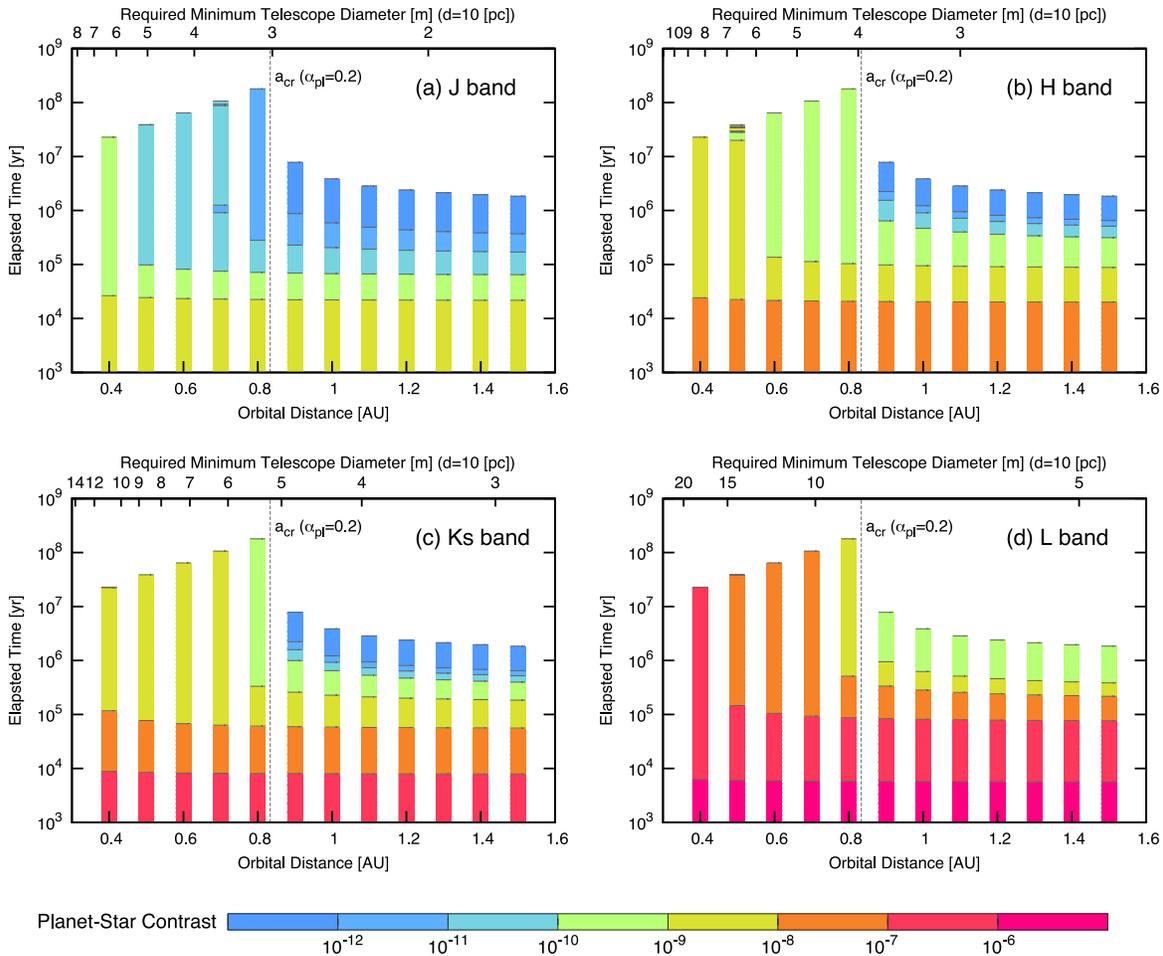}
\caption{Contrast evolution of the thermal emission at each orbital distance with an initial inventory of water of 5 $M_\mathrm{EO}$: (a) J, (b) H, (c) Ks, and (d) L bands. Color shows a planet-star contrast during solidification. The upper horizontal axis in each panel  is the same as Figure \ref{a_contrast}.}
\label{a_elptime}
\end{figure*}

\section{Detectability of magma-covered planets with a steam atmosphere}\label{sec:detect}

Dedicated space coronagraphic missions, such as the Terrestrial Planet Finder C \citep{2009arXiv0911.3200L}, Occulting Ozone Observatory \citep{2010SPIE.7731E..2HS}, WFIRST-AFTA \citep{Spergel15}, and Exo-C and Exo-S\footnote{Exo-C and Exo-S Final Reports in 2015, http://exep.jpl.nasa.gov/stdt/}, have been conducted over long periods of time. The primary goal of such missions is to detect reflected light in visible bands from Earth-sized or super-Earth-sized planets around solar-type stars. These missions will be capable of detecting the reflected light from an Earth-sized planet that is typically $10^{10}$ times fainter than the light of the host star. Moreover, recent progress in extreme adaptive optics has improved the limiting contrast of direct imaging using ground-based telescopes, such as Gemini Planet Imager \citep{2014PNAS..11112661M}, SPHERE \citep{Zurlo14}, Palomar Project 1640 \citep{Oppenheimer12}, and SCExAO \citep{Jovanovic13}. These ground-based high-contrast instruments have an advantage of the small inner working angle on the future 30- to 40-m class telescope. These missions will open the possibility of detecting molten exoplanets. In this section, we discuss the feasibility of the direct detection of molten planets based on the capabilities of these future missions. We also discuss characterization via direct imaging, which would provide evidence of the hot origins of molten planets as well as the basis for more extensive spectroscopic studies for the purpose of characterization of these planets. 

We consider two important quantities that characterize the ease of detection of a planet against stellar speckles. One is planet-to-star contrast, which is defined by the ratio of the planet flux to the stellar flux, and the other is the planet-star angular separation on the celestial sphere. The latter can be converted to the semi-major axis $a$ for a given distance assuming an observer-star-planet angle (phase angle) of $90^\circ$. For future space coronagraphic missions, we regard a contrast of $10^{-10}$ at 0.1 arcseconds of angular separation, which corresponds to an Earth-twin at 10 pc, as a benchmark for detectability. For future ground-based high-contrast instruments, although it is difficult to set a benchmark, we fiducially regard $10^{-7} - 10^{-8}$ of the contrast at 0.01 to 0.03 arcseconds (corresponding to the diffraction limit of the J - L bands for a 30-m telescope), as a benchmark because one of the main goals of these missions is to detect Earth- or super-Earth-sized planets in the habitable zone around late-type stars \citep[see e.g. ][]{2006SPIE.6272E..20M, 2010SPIE.7735E..84M,2012ApJ...758...13K,2012SPIE.8447E..1XG,2013A&A...551A..99C,2014arXiv1407.5099M}.

\subsection{Flux contrast and orbital separation}

Direct detection requires that the planets have a sufficiently high planet-to-star contrast. As described in the previous section, the time evolution of the thermal emission spectra is closely related to the atmospheric evolution during solidification, which strongly depends on the orbital distance from the star. Figure \ref{a_contrast} shows the contrast evolution of the thermal radiation in near-IR windows for various orbital distances in the case of an initial water inventory of 5 $M_\mathrm{EO}$. Although the total contrast consists of the thermal radiation and the reflected stellar light, in most cases, the thermal radiation is dominant in the near-IR band, especially for type-II planets (\S \ref{ss:color}).

The contrast evolution of type-I planets is different from that of type-II planets. For type-I planets, which are located beyond $a_\mathrm{cr}$, the range of the planet-to-star contrast is approximately independent of the semi-major axis. The planet-to-star contrast is initially as high as $10^{-8}$ to $10^{-6}$, and then decreases on a common timescale of less than $\sim$ $10^6$ years (Fig. \ref{a_elptime}). This is because the magma-ocean lifetime of type-I planets is determined by the minimum heat flux, which is approximately independent of the orbital distance. In the case of a smaller inventory of water, the planet-to-star contrast would increase as the atmosphere becomes thinner, whereas the timescale for decreasing planet-to-star contrast becomes shorter. In the case of a larger inventory of water, the thermal radiation reaches the radiation limit earlier. This prolongs the duration of the magma ocean by a factor of at most 2.

The contrasts of type-II planets first decrease with time as well, but then cease to decrease during the course of solidification (Fig. \ref{a_contrast}). The minimum value of the contrast of type-II planets is higher than that of type-I planets. Since the planets are in quasi-radiative energy balance, they spend most of the solidification period at the minimum contrast level in this nominal case (Fig. \ref{a_elptime}). The minimum contrast level is insensitive to the initial water inventory. As long as the planet solidifies on a shorter timescale than the main sequence lifetime of the host star, the minimum contrast level has the highest likelihood of detection. For the case in which the magma ocean is sustained for as long as the main sequence of the star, the planet-to-star contrast gradually increases with time from the minimum contrast level, because the planet must emit a larger planetary radiation from the near-IR band in order to maintain quasi-energy balance in accordance with the luminosity enhancement.

The minimum contrast increases for the planet closest to the star, because this planet receives a larger net stellar flux and so must emit larger planetary radiation. Consequently, planets closer to a star are more favorable for direct imaging in terms of spectral contrast. On the other hand, the planet closest to the star also receives a stronger stellar XUV flux, which fuels hydrodynamic escape. Since the solidification time of type-II planets is approximately equal to the loss time of water, the occurrence rate of molten terrestrial planets is expected to be higher at orbital distances closer to $a_\mathrm{cr}$.

As shown in Figures \ref{typeI&II_evol}, \ref{a_contrast} and \ref{a_elptime}, the type-II molten planets maintain relatively high contrasts in the near-IR band. The type-II planets with $a < 0.5$ AU are favorable targets for future ground-based direct imaging, because the contrasts of the Ks and L bands are above the levels of $\sim 10^{-8}$ and $\sim 10^{-7}$, respectively, for the entire life of the molten state. The duration with the contrast $>10^{-8}$ is approximated with the overall solidification time, which becomes longer with the initial water inventory, as shown in Section \ref{ss:wi}. Although these planets have smaller angular distances, in the near future, a 30- to 40-m ground-based telescope will resolve them within tens of parsecs. On the other hand, the near-IR contrast of the type-I planets declines below $\sim 10^{-8}$ and $\sim 10^{-7}$ on a short timescale of $\sim 10^5$ years for the Ks and L bands, respectively. Young stars should be targeted when searching for type-I planets.  

The thermal radiation in the near-IR band overwhelms the reflected stellar radiation for type-II planets. The contrast of the emission in the visible bands rapidly decreases to the $10^{-10}$ level by a few $10^4$ and $\sim 10^5 $ years for the type-I and type-II planets, respectively (see Figure \ref{typeI&II_evol}).  Therefore, with respect to the visible band, the reflected light should be considered in order to discuss the total contrast of the planets. These issues are discussed in Section \ref{ss:color}. 

\begin{figure}[tbp]
\epsscale{1}
\plotone{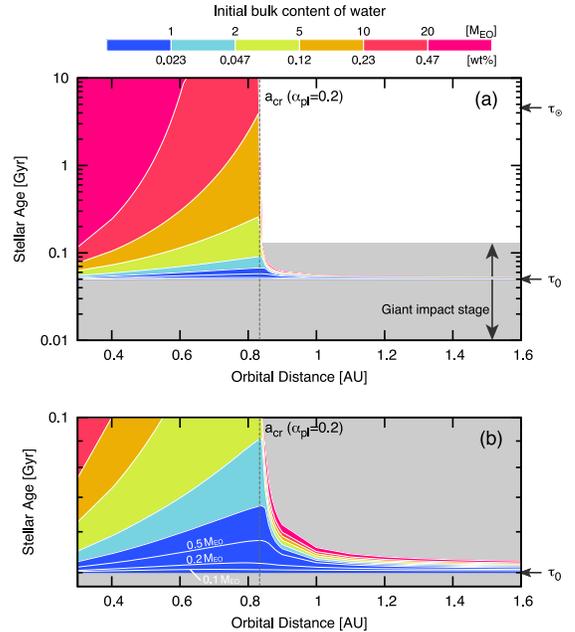}
\caption{Orbital region where molten planets potentially exist as a function of stellar age in a range from 0.01 to 10 Gyr (a), with the assumption that the solidification starts at stellar age $\tau_0$ of 0.05 Gyr. The panel (b) is a zoom-in of the panel (a) for a stellar age younger than 0.1 Gyr. Each color represents the initial bulk content of water required for a molten state to be sustained for a given stellar age and orbital distance. As well as $\tau_0$, the age of the Sun $\tau_\odot$, 4.567 Gyr, is indicated by an arrow on the right side. The critical distance with planetary albedo $\alpha_\mathrm{pl}$ of 0.2 is also shown in each panel.}
\label{MOzone}
\end{figure}

\subsection{Relation between predicted occurrence rate and water endowment\label{ss:wi}}
For the type-II planets, the lifetime of the magma ocean strongly depends on how much water is acquired during planet formation. This suggests the possibility of placing constraints on the initial water inventory based on the observed occurrence rate of hot molten planets. Figure \ref{MOzone} shows an orbital region where molten Earth-sized planets potentially exist as a function of stellar age. The predicted lifetime of the magma-covered planets shows a sharp transition around $a_\mathrm{cr}$. The type-I planets during the magma ocean period have less probability of being detected, whereas the probability of detecting molten type-II planets is greatly enhanced with initial water inventory.

The calculated lifetimes of type-I and type-II planets allow us to estimate expected number of G-type stars which host Earth-sized molten planets. For G-type stars whose distance from the solar system is less than the distance $d$, the expected number of G-type stars with molten planets $N_\mathrm{D} (d)$ can be estimated as follows, 
\begin{eqnarray}
N_\mathrm{D}(d) &\sim& N_\mathrm{G}(d)\cdot P_\mathrm{EG}\cdot \frac{\Delta t_\mathrm{mol}^\mathrm{tot}}{\bar{\tau}_\mathrm{G}(d)},
\end{eqnarray}
where $N_\mathrm{G}(d)$ and $\bar{\tau}_\mathrm{G}(d)$ are, respectively, the number and average age of G-type stars within the distance $d$ from the solar system, $P_\mathrm{EG}$ is the fraction of G-type stars with terrestrial planets in a range of planetary orbital distances considered, $\Delta t_\mathrm{mol}^\mathrm{tot}$ is the total duration during which terrestrial planets are in a molten state. Let us consider the case when $d$ is 10 pc. The total number of G-type stars is then around 30. The average age of G-type stars is assumed to be 5 Gyr. Using Kepler data, \citet{Petigura13} reported that the fraction of GK-type stars which host Earth-sized planets is $\sim$ 0.12 for orbital distances of 0.05-0.42 AU, and $\sim$ 0.086 in the habitable zone. Here, we adopt $P_\mathrm{EG}$ of 0.1 for both type-I and -II planets. 

Type-I planets solidify on a timescale of the order of $10^6$ years, which shorter than typical time intervals of giant impacts at late stage of formation. In this case, the total duration would be expressed as a product of an average number of giant impacts $n_\mathrm{GI}$ and the duration of a molten state after one collision $\Delta t_\mathrm{mol}$. According to N-body simulations, Earth-like planets would experience 10-20 giant impacts during formation \citep{KG10, SL12}. We adopt $n_\mathrm{GI}$ of 10. Using a typical duration of a magma ocean of $2\times10^6$ years, we obtain $\sim$ 0.01 for $N_\mathrm{D}(d=10\, \mathrm{pc})$ in the type-I orbital region. This value is common for type-I planets which formed with initial water inventory comparable to or exceeding the total amount of water of the modern Earth. The potentially favorable target for molten type-I planets would be young and other spectral type stars such as Fomalhaut. On the other hand, type II planets have a wide range of solidification time after single impact, which strongly depends on the initial amount of water. In the extreme case that the solidification time exceeds the period of giant impact phase, which probably lasts until the stellar age of 0.1 Gyr, the total duration of a molten state is given by the duration of a molten state after single collision. In the type-II orbital region, the initial amount of water thus greatly affects the probability of detecting molten planets. Given that $\Delta t_\mathrm{mol}^\mathrm{tot}$ does not exceed 10 Gyr, $N_\mathrm{D}(d=10\, \mathrm{pc})$ ranges from $\sim$ 0.01 to 6 in the case of type-II planets, considering the initial inventory of water of from one-tenth to several tens times the Earth's ocean mass.  

Here, we address the uncertainties in parameters which affect the loss time of water, the timing of the last giant impact, the heating efficiency and the EUV flux from the G-type host star. The giant impact phase is triggered by the dispersal of nebula gas and lasts until a stellar age of approximately 0.1 Gyr. Therefore, the last giant impact would occur between the stellar ages of 0.01 and 0.1 Gyr. The heating efficiency $\eta$ in Eq. (\ref{eq:loss_rate}) depends on the structure and details of the energy budget in the upper atmosphere and remains poorly constrained. The estimated values have a large spread and typically range from 0.1 to 0.5 \citep[e.g.][]{Watson81, Chassefiere96, Kulikov07, Shematovich14}. The saturation level of XUV flux and its ending time $\tau_\mathrm{sat}$ remain quite uncertain. If we simply assume that the XUV-to-bolometric-luminosity ratio scales as the X-ray-to-bolometric-luminosity ratio, which is observationally estimated to be $10^{-3.2\pm 0.3}$ during the saturation phase for one solar-mass star \citep{Pizzolato03}, the variation of the XUV saturation level is nothing more than a factor of 3 and $\tau_\mathrm{sat}$ ranges from 0.06 to 0.13 Gyr. Although the XUV flux could exhibit greater fluctuation for pre-main sequence stars, its effects might not be so significant because of its short duration relative to the subsequent evolution. Although no observational data are available at wavelengths between 360 and 921 $\mathrm{\AA}$ due to strong absorption by the interstellar medium, the total XUV flux would not have profound uncertainty compared with that of the other parameters, because the estimated energy flux in this wavelength range is at most 10 to 20\% \citep{Ribas05}.

Although our model cannot predict the exact inventory of primordial water due to the numerous uncertainties described above, estimating the order of magnitude of the exact inventory of primordial water would be still useful. Our model suggests that, if the initial bulk content of water exceeds approximately 1 wt\%, molten planets would be present over the main sequence of the host G-type star (Fig. \ref{MOzone}). Since the critical distance, which separates the type-I and type-II planets, originates from the presence of the radiation limit of steam atmospheres, detection of a critical distance would offer a diagnostic test to determine whether water is generally one of the major volatile species of extrasolar terrestrial planets. In the orbital region of type-II planets, the detection of hot signatures in the planetary spectrum will provide a lower limit for initial water inventory, which is sufficient to sustain the molten state for a given stellar age, whereas no detection will constrain its upper limit, so that the planet will lose most of its water and become solidified.

\begin{figure}[tbp]
\epsscale{1}
\plotone{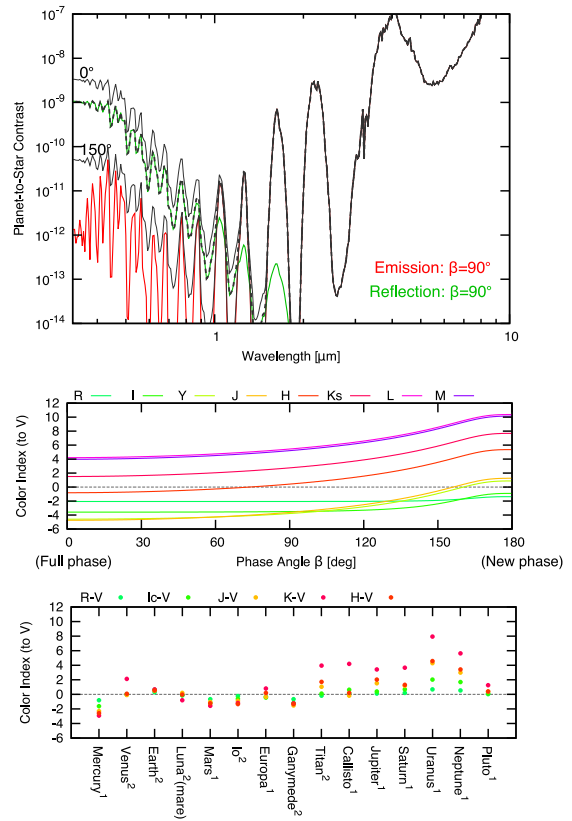}
\caption{Superimposed spectra and band color of a planet at 0.6 AU with an atmospheric pressure of 100 bar and a surface temperature of 2,000 K. Under this condition, the planet is in quasi-energy balance. The thin black lines in the top panel indicate the planet-to-star contrast of superimposed spectra. The spectrum components are indicated by the red line for the thermal emission and the green line for the reflected light, respectively, for a phase angle of 90$^\circ$. The middle panel shows the color index of each band to the V band relative to the host star. As a reference, color indexes relative to the Sun are plotted in the bottom panel for all planets \citep{Lundock09}. The superscript attached to the planet's name denotes the observational period of the data used: 1 for May 2008 and 2 for 2008 Nov.}
\label{phase}
\end{figure}

\subsection{Reflected light and color variations of solidifying planets \label{ss:color}}
The actual spectrum from the planet is obtained as the superposition of its thermal emission and reflected stellar light. The thermal radiation generally overwhelms the reflected stellar radiation during the highly self-luminous stage. Figure \ref{phase} shows the planet-to-star contrast of the superimposed spectra of the planet in quasi-energy balance. According to the Lambert assumption, the planet-star contrast $C_\mathrm{ps}^\mathrm{ref}(\lambda)$ of the reflected light is given as follows:
\begin{eqnarray}
  \label{eq:reflection}
	C_\mathrm{ps}^\mathrm{ref}(\lambda) &=& \frac{2\phi(\beta)}{3} \alpha(\lambda) \left( \frac{R_\mathrm{pl}}{a} \right)^2 \\
	\phi(\beta) &=& \frac{\left[  \sin \beta + (\pi-\beta)\cos\beta \right]}{\pi},
\end{eqnarray}
where $\alpha$ is the spectral Bond albedo, and $\phi(\beta)$ is the Lambert phase function with respect to the phase angle $\beta = \angle (\mathrm{star-planet-observer})$. The overall superposed spectrum in the quasi-energy balance state is V-shaped with a bottom around 1  $\mathrm{\mu m}$, except for the phase angle very close to the new phase (Fig. \ref{phase}). The dominant component is thermal emission at wavelengths longer than $\sim$ 1 $\mathrm{\mu m}$ and is reflected stellar light at the shorter wavelengths. This originates from the difference in the peak wavelength of the Planck function between the surface temperature and the stellar effective temperature. The peak wavelengths are approximately 1.4 $\mathrm{\mu m}$ for 2,000 K and 500 nm for 5,800 K.

\begin{figure}[tbp]
\epsscale{1}
\plotone{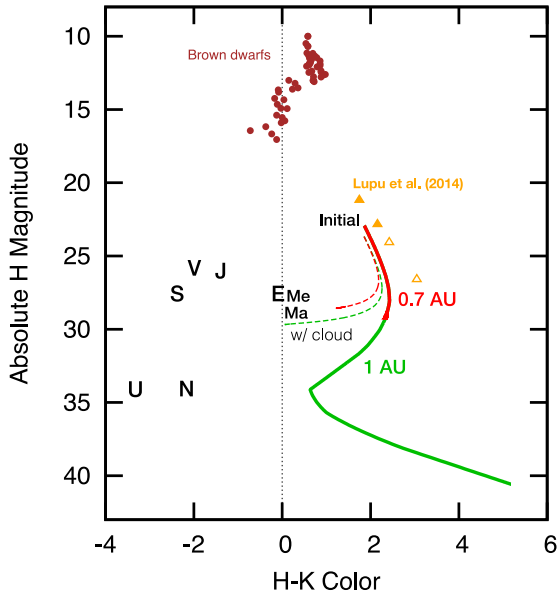}
\caption{Color-magnitude relation of the thermal emission from the solidifying planets with a steam atmosphere at 1 AU (green) and 0.7 AU (red) with an initial water inventory of $5 M_\mathrm{EO}$. The dashed lines indicate data for the cases in which the stellar light reflected by clouds is taken into account with a phase angle of 90$^\circ$. Here, we assume that the clouds are uniformly distributed and the coverage and albedo are 0.5. We adopt the H-K color of the Sun (0.05) as that of the host star. The brown circles are observed brown dwarfs \citep{Leggett02, Knapp04}. The letters represent the planets in the solar system \citep{Lundock09}. The results reported by \citet{Lupu14} are also shown for total surface pressures of 10 bar (filled triangles) and 100 bar (open triangles) for surface temperatures of 2,200 and 1,600 K for the case in which the atmospheric composition is in equilibrium with magma having the composition of the Earth's continental crust.}
\label{color-mag}
\end{figure}

For $\beta=90^\circ$, the contrast of the reflection can be estimated as follows:
\begin{eqnarray}
  \label{eq:reflectionorder}
	C_\mathrm{ps}^\mathrm{ref} \approx 3 \times 10^{-10} \left( \frac{\alpha(\lambda)}{1} \right)  \left( \frac{R_\mathrm{pl}}{R_\oplus} \right)^2   \left( \frac{a}{1 \mathrm{AU}} \right)^{-2},
\end{eqnarray}
where $R_\oplus$ is the Earth radius. Because the albedo in the U-B band is $\sim 1$, as shown in Figure \ref{fig:LBLspec}(b) and the reflection dominates in this band, we roughly expect contrasts of $\sim 10^{-10}$ and $\sim 10^{-9}$ for the type-I and type-II planets, respectively. These contrasts are within the scope of the future space direct imaging survey. The contrasts decline steeply from the visible wavelengths to the near-IR wavelengths, which reflects the deep atmosphere of the planets. 

The reflected light is bluer than that of the host star, whereas the thermal emission in the near-IR is redder than that of the host star. The middle panel in Fig. \ref{phase} shows the color index of each band to the V band relative to the host star as a function of phase angle. As the phase angle increases, the color becomes redder for all bands because the contribution of the thermal emission increases. The color indexes of the I and R bands remain bluer than that of the host star, whereas the color indexes of the other spectral bands becomes redder than that of the host star. The color indexes could vary by approximately 6 for the J, H, Ks, and L bands during a complete cycle of orbital motion. These color changes are comparable to the color differences between rocky planets, such as Earth and Mars, and gaseous planets such as Uranus and Neptune.

Following \citet{Lupu14}, we consider the color-magnitude diagram of the molten planets. Figure \ref{color-mag} shows the evolutionary paths of the thermal emission in the H-K color and absolute H magnitude diagram. The molten planets are as faint as the planets in the solar system, but are distinctly redder in color. This is because the thermal emission, which is much redder than the stellar light, dominates the planetary spectrum for molten planets, whereas the color of the planets in the solar system mostly reflects the reflected light. We also estimate the effects of stellar light reflected by clouds on the H-K color by assuming that the cloud coverage is 0.5. In this case, as the thermal emission becomes fainter, the reflected light becomes more dominant in the near-IR spectrum. The reflected light could determine the color at the later stage for the planet at 1 AU, whereas the color is still much redder than that of the star during solidification of the planet at 0.7 AU, including the quasi-energy balance state. A self-consistent evolutionary calculation, including cloud formation, will be the subject of future research to examine actual color variations that occur during the course of solidification.

\section{Discussions}\label{sec:discuss}
\begin{figure}
\epsscale{1.1}
\plotone{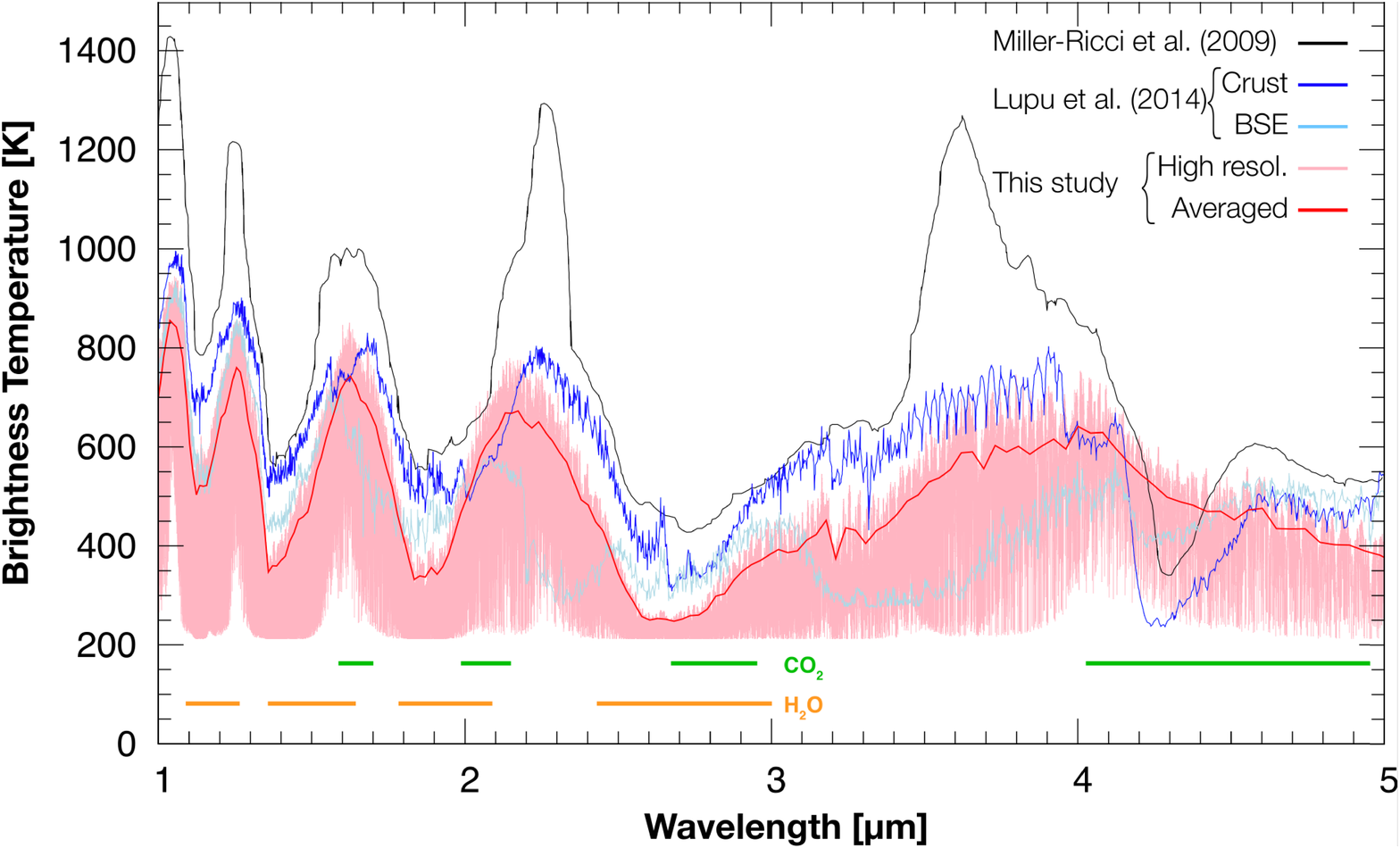}
\caption{Brightness temperatures obtained by the different models. The total atmospheric pressure is 10 bar in all models, whereas the surface temperature is 1,500 K for \citet{M-R09} and 1,600 K for \citet{Lupu14} and our model. We consider a pure steam atmosphere. The model atmosphere consists of 90\% $\mathrm{H_2O}$ and 10\% $\mathrm{CO_2}$ for \citet{M-R09}. \citet{Lupu14} considered atmospheres in equilibrium with the hot surface having the composition of the Earth's continental crust and bulk silicate Earth (BSE). Although these model atmospheres are composed by various gaseous species, they consist primarily of $\mathrm{H_2O}$ and $\mathrm{CO_2}$. \citet{Lupu14} stated that, by taking into account quenching effects, the spectra for BSE composition magmas become similar to that of the Earth's continental crust.}
\label{Teff_comp}
\end{figure}

We compare the brightness temperatures of our model with those reported by \citet{M-R09} and \citet{Lupu14} for wavelengths ranging from 1 to 5 $\mathrm{\mu m}$ (Fig. \ref{Teff_comp}). Although the atmospheric compositions differ among these models, our model spectrum is more similar to that in \citet{Lupu14}, especially for the atmospheric composition in a chemical equilibrium with silicate melts having the composition of the Earth's continental crust. The brightness temperature, i.e., the thermal radiation, of the our model tends to be lower than that of \citet{Lupu14} for continental-crust-composition magmas, probably because of the $\mathrm{H_2O}$-rich composition and the lower temperature in the upper atmosphere of our model. 

Although we have thus far discussed only pure steam atmospheres, the inclusion of other gaseous species would not substantially alter our conclusions. Adding other opacity sources would reduce the planet-to-star contrasts in atmospheric windows for type-I and -II planets in a self-luminous state, whereas doing so would only affect the relative contrast of atmospheric windows for the type-II planets in quasi-energy balance. If the thermal radiation decreases at certain wavelengths during the quasi-energy balance state, the planet must emit larger thermal radiation at other wavelengths in order to maintain the energy balance while reducing the amount of the steam atmosphere. The same would be said of the effects of clouds. At very early stage, a molten planet would have no cloud cover because of the hot temperature. As the temperature drops and the atmosphere grows, however, clouds would start to form by condensation of water vapor in the atmosphere. Clouds would increase the planetary albedo, which makes the critical distance smaller by reducing the net incident stellar radiation. Clouds also would enhance a greenhouse effect, which makes the critical distance larger. The presence of clouds therefore would affect the actual position of the critical distance through a balance of the opposing effects of cloud albedo and a greenhouse effect. However, as long as a planet is type II, it would emit strong thermal radiation through the atmospheric windows even with some cloud cover, to maintain energy balance for the same reason as the case with additional gaseous opacity described above.

To a first order approximation, we have assumed that the planetary interior has an adiabatic temperature profile in our evolutionary model. Although horizontal inhomogeneity would exist in the convective magma ocean, especially around boundary layers, the planet would have a nearly uniform surface temperature under a thick and massive steam atmosphere considered in the present study, because the timescale of heat transfer by atmospheric circulation is generally much shorter than the timescale of radiative cooling and of heat transfer by convection of the magma ocean.

In the evolutionary model, we have also assumed that the magma ocean effectively transports internal heat to the surface and that the atmosphere is in solution equilibrium with the magma ocean for calculations of a degassing rate. Since the assumption of the solution equilibrium leads to a maximum estimate of the atmospheric growth rate, the thermal radiation predicted in the present study would be a minimum estimate, whereas the decline timescale of the emergent spectra would be overestimated.

Heat flux and a degassing rate from a magma ocean strongly depends on the magma viscosity. Thermal structure and heat transfer of a magma ocean has been extensively reviewed by \citet{Solomatov07}. Viscosity of ultramafic silicate melt near liquidus is as low as 0.1 Pa s at low pressures \citep[e.g.][]{Shaw72}. The similar value can be assumed for the magma viscosity under the mantle conditions because of small temperature and pressure effects for a low-viscosity and completely depolymerized melt (see discussions by \citet{Solomatov07}). As the Rayleigh numbers are typically in the range of $10^{20}$-$10^{30}$ with the low viscosity \citep{Solomatov07, E-T08}, the magma ocean at early stage is expected to be highly turbulent and in a state of vigorous convection. The highly turbulent magma ocean is considered to be homogeneous and its thermal structure is to be nearly adiabatic. 

Along with cooling, the magma ocean starts to solidify at great depth, because the slopes of liquidus and solidus of Earth's mantle are steeper than those of adiabats in the magma ocean \citep{Andrault11}. On the other hand, when a solid crust starts to form depends on the energy balance at the surface. The energy balance  between heat flux from the magma ocean and radiative heat flux from the surface determines the surface temperature. A steam atmosphere has a strong blanketing and greenhouse effect. Even with a surface atmospheric pressure of 10 bar, its total opacity exceeds $10^5$ in the 10-$\mathrm{\mu m}$ atmospheric window. Therefore, the insulating atmosphere maintains the surface in a molten state if the heat flux from the magma ocean is sufficiently high. 

The most important parameter for heat transport of the magma ocean is the magma viscosity, which varies dramatically depending on the melt fraction in the magma. Experimental studies have shown that viscosity of partial melts abruptly increases from a liquid-like low value to a solid-like high value (rheological transition), when its melt fraction falls below a critical value, approximately 40 \% \citep[e.g.][]{MP79}. As long as the surface temperature is high enough to keep the melt fraction larger than the critical value, the magma ocean can transport the internal heat to the surface efficiently by vigorous convection, and also keep a high renewal rate of the surface. As the surface temperature decreases, the melt fraction decreases and reaches the critical value, and then solid-state convection takes over. The heat flux from the magma ocean is no longer high enough to keep the surface molten. The surface temperature rapidly drops and a conductive lid crust starts to develop at the surface \citep{Solomatov07, Lebrun13}. 

As the solidification proceeds, the depth of the magma ocean becomes shallow and the total mass of the magma decreases. The magma ocean becomes enriched in water, which is incompatible to silicate cumulates. The solution equilibrium of water between the atmosphere and the magma ocean is assumed in our model, whereas actually a certain degree of supersaturation of water might be required for degassing by bubbling and $\mathrm{H_2O}$ diffusion processes. \citet{Hamano13} discussed the effects by the simple treatment for the degassing processes on the overall solidification time of the magma ocean. For degassing involving the bubble formation, bubbles are required to grow fast and detach from the magma flows.  Although the supersaturation required for the efficient bubble detachment remains uncertain, even using 50-fold water solubility shortens the solidification time by a factor of at most 4, based on the results by \citet{Hamano13}. They also estimated the supersaturation required for the replenishment of the atmosphere against its loss by the molecular diffusion process only, using a molecular diffusion coefficient in basaltic melt by \citet{Zhang07}. They showed that a modest degree of supersaturation is enough for the replenishment, as long as the magma viscosity is a liquid-like low value. Eddy diffusion process would significantly lower the required supersaturation. However, at a very late stage after the melt fraction reaches the critical value, both the resurfacing rate and the efficiency of bubble detachment from the magma flows would be greatly reduced due to the solid-like high viscosity. Redefining the solidification time as the time required for the surface temperature to drop to 1,700 K, around which the melt fraction is estimated to reach the critical fraction for the rheological transition, shortens the lifetime of the magma ocean on type-I planets by a factor of less than 5 in the parameter range considered herein. This estimate is consistent with the typical duration of the magma ocean phase of 1 Myr reported by \citet{Lebrun13}. For type-II planets, the redefinition of the solidification time has less effect on the lifetime of the magma ocean, as long as the water endowment is sufficiently large so that the quasi-energy balance is attained at a surface temperature above 1,700 K. These uncertainties would be smaller compared to the parameter uncertainties associated with the energy-limited escape rate, as discussed above.

Although we have examined only Earth-sized planets in the present study, the orbital-dependent dichotomy in lifetime and spectral evolution would exist for different planetary masses as well. \citet{Kopparapu14} reported that the radiation limit increases by approximately 20 \% for a rocky planet with 5 times the Earth's mass. The critical distance therefore would become closer to the host star for larger planetary masses, compared to that for Earth-sized planets. Although the larger radiation limit also would have a role to make the cooling of a type-I planet faster, the overall duration of a molten state may be roughly proportional to the planetary mass because of the larger heat capacity of the planet \citep{E-T11}. On the other hand, for a massive type II planet, the lifetime of a magma ocean would become extremely longer, with the assumption that the bulk content of water is the same. This is because the total amount of water is proportional to the planetary mass, whereas the total escape rate of hydrogen is independent on the planetary mass itself, as long as the energy-limited escape rate is considered. In terms of spectral detectability, more massive planets are expected to be brighter and therefore easier to detect because of their larger surface area \citep{M-R09}. Future studies will expand this line of research by taking into account planetary mass, different types of stars and an expected variety of atmospheric compositions.

\section{Conclusions}\label{sec:conclusion}
We have examined the lifetime of the magma ocean and its spectral evolution for Earth-sized terrestrial planets along with the evolution of the steam atmosphere. Due to the presence of the radiation limit of steam atmospheres, the thermal and spectral evolution of the magma ocean strongly depends on the planetary orbital distance from the host star. Planets beyond the critical distance $a_\mathrm{cr}$ (type I) are self-luminous throughout solidification. The thermal radiation from atmospheric windows decreases on a timescale shorter than $\sim 10^6$ years. The lifetime of the magma ocean is less sensitive to the semi-major axis in the type-I orbital region. Consequently, young stars should be targets in searching for molten planets in this orbital region.

In the case of a planet formed inside $a_\mathrm{cr}$ (type II), the outgoing planetary radiation can balance the net incoming stellar radiation after a short self-luminous phase. The emergent spectra during the quasi-energy balance state are approximately independent of the surface temperature and pressure. The type-II planet maintains high thermal radiation from near-IR atmospheric windows throughout the lifetime of the magma ocean. The lifetime of the magma ocean increases for a larger initial inventory of water.

Type-II planets are brighter at smaller orbital distance so as to balance the larger stellar radiation, whereas these planets solidify on a shorter timescale because of the higher loss rate of water they experience. The sensitivity of the magma-ocean lifetime to the water amount and the orbital distance increases the possibility of placing constraints on the water endowment of terrestrial planets based on the occurrence rate of molten planets for a given stellar age. Our model predicts that terrestrial planets remain molten over the main sequence of the host star in the type-II orbital region if the initial bulk content of water exceeds approximately 1 wt\%. 

The composite spectrum is dominated by reflected light at the visible wavelengths and by thermal emission at the near-IR wavelengths for type-II planets in the quasi-energy balance state. The atmospheric windows in the near-IR are favorable for detecting thermal emission, as pointed out by \citet{Lupu14}. The H-K color of the thermal emission is strikingly redder than that of the host star and the reflected light, whereas the solidifying type-I and type-II planets are typically as faint as planets having moderate surface temperature. Therefore, type-I and type-II planets could form a group distinct from the observed brown dwarfs and the planets in the solar system in the color-magnitude diagram. Evolution calculations including cloud formation will be required in order to make more detailed predictions regarding color and appearance.

The K and L bands will be favorable for future direct imaging because the planet-to-star contrast of the type-II planets is approximately $10^{-7}-10^{-8}$. In particular, the search around $a_\mathrm{cr}$, which is close to the inner edge of the habitable zone, will have a higher detection probability in terms of the duration of the molten state. For both the type-I and type-II planets, in visible atmospheric windows, the contrast of the thermal emission drops below $10^{-10}$ in less than $10^{5}$ years, whereas that of the reflected stellar light remains $10^{-10}$. Thus, the visible reflected light from molten planets will also be a promising target for direct imaging with ground-based high-contrast instruments and for second-Earths survey by space-based telescopes.  
 
 \acknowledgments
 We are grateful to Colin Goldblatt for allowing us to use his spectral data for model comparison. We appreciate two anonymous reviewers for their careful reading and constructive comments. This work is supported by Grant-in-Aid for Scientific Research on Innovative Areas from the Ministry of Education, Culture, Sports, Science and Technology (MEXT) (No. 23103003). This work is also supported by Grant-in-Aid for Young Scientists (B) from Japan Society for the Promotion of Science (JSPS), No. 26800242 to K.H., No. 25800106 to H.K., and No. 25800264 to M.O. H.K. acknowledges support from Research Center for the Early Universe (RESCEU).
 

\appendix
\begin{figure}[bp]
\epsscale{0.6}
\plotone{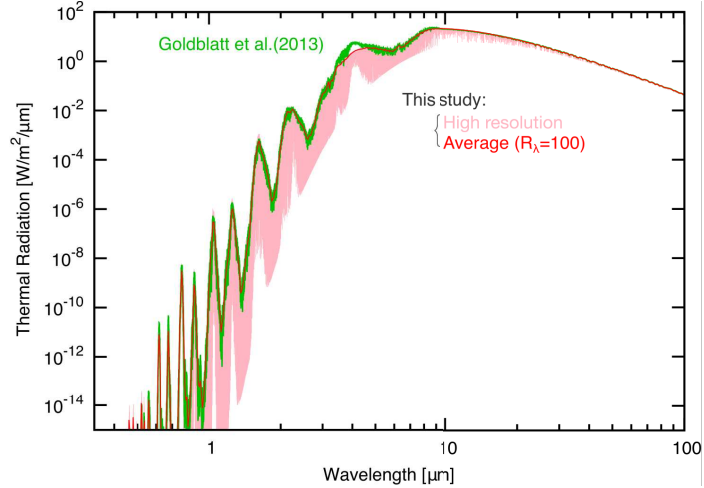}
\caption{Spectrum of the thermal emission from the top of the steam atmosphere obtained using our model and that reported by \citet{Goldblatt13} for a surface pressure of 123 bar and a surface temperature of 600 K.}
\label{fig:spec_comp}
\end{figure}

\section*{Radiative Transfer}
We use two-stream approximation to solve radiative transfer equation in a plane-parallel atmosphere, with diffusivity factor $r_\mathrm{df}$ of $\sqrt{3}$,
\begin{eqnarray}
	r_\mathrm{df}\frac{dF^{\uparrow}_\nu}{d \tau} &=& F^{\uparrow}_\nu - J_\nu(\tau) \nonumber\\
	- r_\mathrm{df}\frac{dF^{\downarrow}_\nu}{d \tau} &=& F^{\downarrow}_\nu  - J_\nu(\tau).
	\label{eq:RT}
\end{eqnarray}
Hereafter, the wavenumber subscript $\nu$ is suppressed to simplify the notation. We assume that scattering is isotropic, and treat stellar and planetary radiation separately. For the planetary flux ($F^{\uparrow}_\mathrm{pl}$ and $F^{\downarrow}_\mathrm{pl}$) , 
\begin{eqnarray}
	J(\tau) = J_\mathrm{pl}(\tau) = \frac{\omega_0}{2}\left(F^{\uparrow}_\mathrm{pl}+F^{\downarrow}_\mathrm{pl} \right) + (1-\omega_0)\pi B\left(T(\tau)\right),
	\label{eq:Jpl}
\end{eqnarray}
where $\omega_0$ is the single scattering albedo, and $B\left(T(\tau)\right)$ is the Planck function with temperature at $\tau$. Boundary conditions at the top and bottom of the atmosphere are given by
 \begin{equation}
	F^\downarrow_\mathrm{pl,t} = 0\;\;\; \mathrm{and} \;\;\;
	F^\uparrow_\mathrm{pl, b} = (1-A_\mathrm{g}) \pi B(T_\mathrm{s}) + A_\mathrm{g} F^\downarrow_\mathrm{pl, b},
	\label{eq:BCpl}
\end{equation}
where $A_\mathrm{g}$ is the ground albedo and $T_\mathrm{s}$ is the surface temperature. The subscripts $\mathrm{t\ and\ b}$ denote the top and the bottom of the atmosphere, respectively.

For the stellar diffusive flux ($F^{\uparrow}_\mathrm{sl}$ and $F^{\downarrow}_\mathrm{sl}$) ,  
\begin{eqnarray}
	J(\tau) = J_\mathrm{sl}(\tau) = \frac{\omega_0}{2}\left(F^{\uparrow}_\mathrm{sl}+F^{\downarrow}_\mathrm{sl} \right) + \frac{r_\mathrm{df}\omega_0}{2}D_\mathrm{\odot} e^{\tau/\mu_0},
	\label{eq:Jsl}
\end{eqnarray}
where $D_\odot$ is the stellar flux incident to the top atmospheric layer at optical depth of $\tau_t$ with zenith angle $\theta_0$ ($\mu_0 = \cos \theta_0$). The direct downward stellar flux is given by 
\begin{eqnarray}
	D^{\downarrow}_\mathrm{sl}(\tau) &=& |\mu_0| D_\odot e^{\tau/\mu_0}.
	\label{eq:Dsl}
\end{eqnarray}
We assume that the stellar spectrum profile incident to the top of the atmosphere is given by that of a blackbody radiation with an effective stellar temperature $T_\mathrm{\bigstar}$. It gives the boundary condition as follows,
\begin{equation}
	F^\downarrow_\mathrm{sl, t} = 0,\;\;\; 
	D_\odot = F_\odot \frac{\pi B(T_{\bigstar})}{\sigma T_{\bigstar}^4},\;\;\;  \mathrm{and} \;\;\;
	F^\uparrow_\mathrm{sl, b} = A_\mathrm{g} \left( F^\downarrow_\mathrm{sl, b}  + D_\mathrm{sl, b}^{\downarrow} \right),
	\label{eq:BCsl}
\end{equation}
where $F_\odot$ is the total stellar energy flux that the planet receives at its orbital position.

Wavenumber region is separated into two parts to calculate the radiative transfer equation efficiently and accurately. In the wavenumber region less than a certain wavenumber $\nu_\mathrm{a}$, we assume that the scattering process by gaseous molecules can be neglected. The net upward planetary and stellar fluxes are calculated using the integrated form of Eq. (\ref{eq:RT}), using Eqs. (\ref{eq:Jpl}), (\ref{eq:Jsl}) and (\ref{eq:Dsl}), and the boundary conditions (\ref{eq:BCpl}) and (\ref{eq:BCsl}),
\begin{eqnarray}
	F^\uparrow_\mathrm{pl} (\tau) &=& \pi B(T(\tau)) 
	+ \pi \int_{\tau}^{\tau_\mathrm{b}} \frac{d  B(T(t))}{dt} e^{-r_\mathrm{df}|t-\tau|} dt 
	+ \left\{  F^\uparrow_\mathrm{pl, b}  - \pi B\left( T (\tau_b)\right)  \right\} e^{-r_\mathrm{df}(\tau_\mathrm{b}-\tau)} \nonumber \\
	F^\downarrow_\mathrm{pl} (\tau) &=& \pi B(T(\tau))  
		 - \pi \int_{0}^{\tau} \frac{d  B(T(t))}{dt} e^{-r_\mathrm{df}|t-\tau|} dt
		 -  \pi B\left( T (0) \right)e^{-r_\mathrm{df}\tau}
	 \label{eq:intRTpl}
\end{eqnarray}
and
\begin{eqnarray}
	F^\uparrow_\mathrm{sl}(\tau) &=& F^\uparrow_\mathrm{sl, b} e^{-(\tau_\mathrm{b}-\tau)} = A_\mathrm{g} e^{\tau_b/\mu_0}e^{- (\tau_\mathrm{b} - \tau)} \nonumber \\
	F^\downarrow_\mathrm{sl}(\tau) &=& 0,
\end{eqnarray}
given the fact that the diffusive downward stellar flux is equal to 0 in a non-scattering atmosphere. We evaluate the integrals in Eqs (\ref{eq:intRTpl}) by considering the contributions from the region where the optical distance from the midpoint of each layer is smaller than 10. 

At sufficiently large wavenumber, both the absorption and scattering processes are taken into account. The radiative flux in each layer is calculated using a formal solution of the radiative transfer equation (\ref{eq:RT}) similarly to \citet{Toon89}, but is solved iteratively by repeating upward and downward integrations to satisfy the boundary conditions at the top and the surface.

\begin{figure*}[tbp]
\epsscale{1}
\plottwo{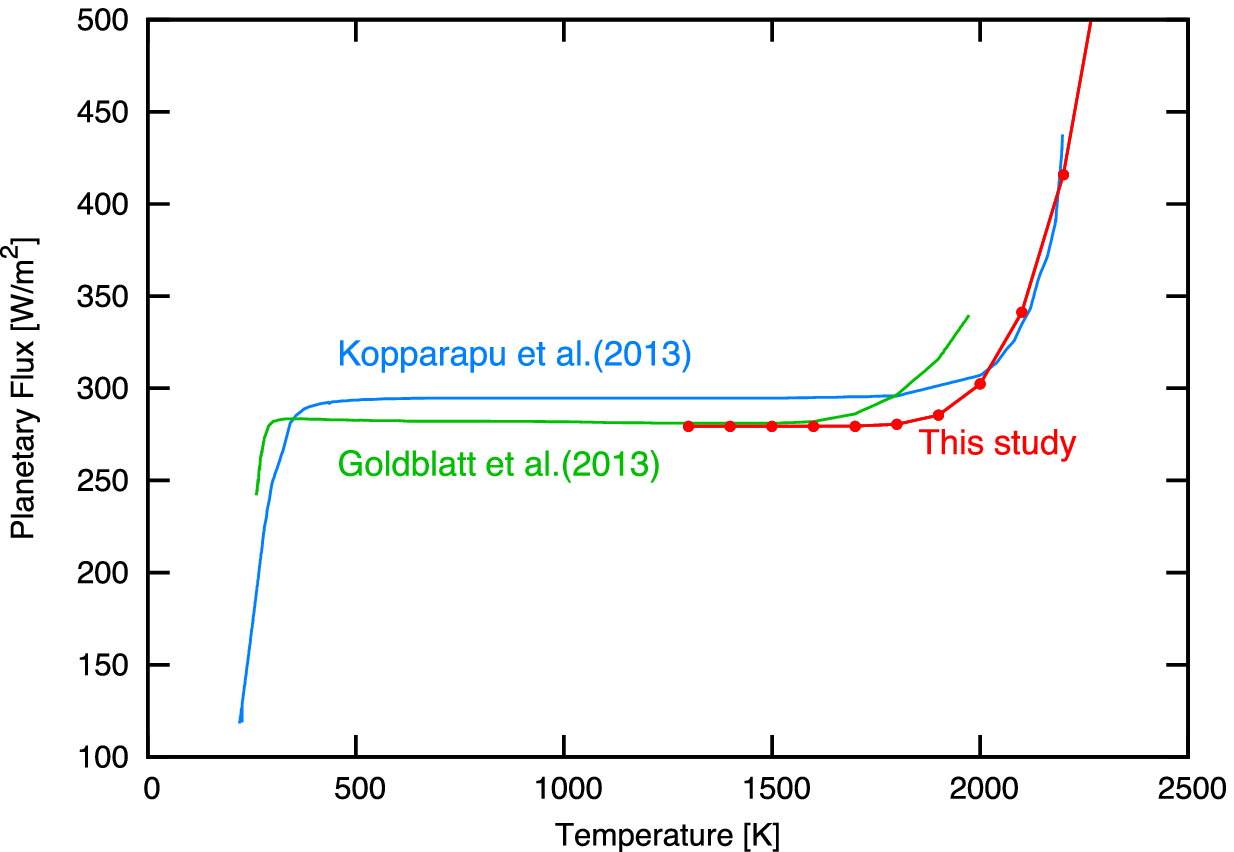}{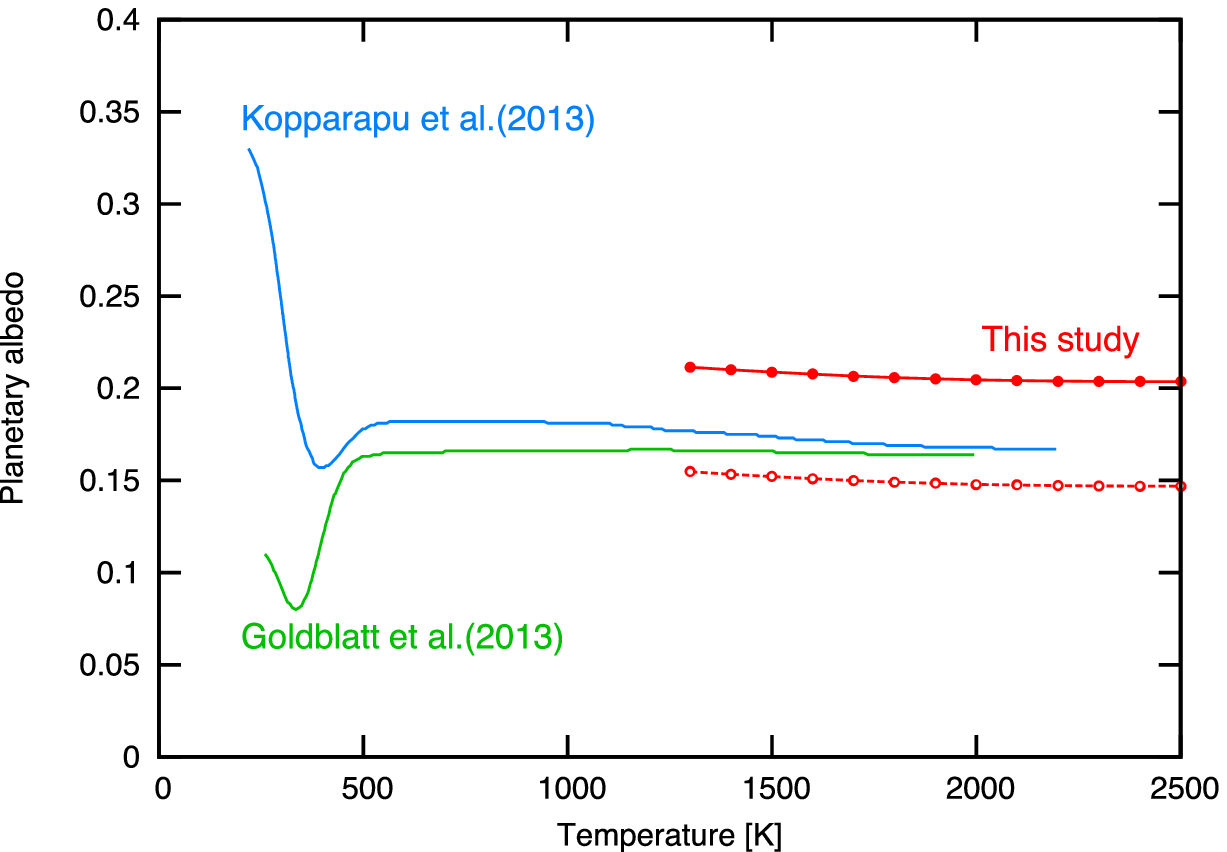}
\caption{Planetary radiation and Bond albedos for a steam atmosphere with a surface pressure of 260 bar. \citet{Kopparapu13} and \citet{Goldblatt13} obtained the Bond albedos by averaging the stellar flux over six zenith angles, whereas we calculated the Bond albedos using a stellar zenith angle of $\cos^{-1} ( 1/\sqrt{3} )$ (red closed circles and solid line). The open circles and dotted line represent the Bond albedos obtained by assuming that the stellar radiation is perfectly absorbed at wavenumbers beyond 30,000 $\mathrm{cm^{-1}}$.}
\label{fig:Fpl_comp}
\end{figure*}

In this paper, we calculate the radiative transfer equations for both the planetary and stellar radiation in a wavenumber range from 0 to 30,000 $\mathrm{cm^{-1}}$. The wavenumber $\nu_a$, at which the way to calculate the radiative transfer equation is switched from one to another, is set to 5,000 $\mathrm{cm^{-1}}$. Beyond 30,000 $\mathrm{cm^{-1}}$, only the scattering process is considered using the Rayleigh scattering cross-section at 30,000 $\mathrm{cm^{-1}}$. This leads to underestimate the attenuations by the absorption and scattering processes, and overestimate the Bond albedo (see Fig. \ref{fig:spec_comp}). We define a spectral albedo $\alpha$ for the incident stellar radiation as the ratio of the diffusive upward radiation to the direct downward radiation at the top of the atmosphere, i.e., $F^\uparrow_\mathrm{sl} (\tau_\mathrm{t}) / D^\downarrow_\mathrm{sl, t}$. The Bond albedo for stellar radiation with an effective temperature of $T_\bigstar$ is obtained as a weighted mean of the spectral albedo as follows,
\begin{eqnarray}
	\alpha_\mathrm{p} &=& \frac{\int \alpha B (T_\bigstar) d \nu}{\sigma T_\bigstar^4}.
\end{eqnarray}

We compare the thermal emission spectra of our model with that reported by \citet{Goldblatt13} (Fig. \ref{fig:spec_comp}). Our model spectra accord closely with that by \citet{Goldblatt13}. The thermal radiation calculated with our model is lower than that by \citet{Goldblatt13} at the wavelengths between 3 and 5 $\mathrm{\mu m}$, due to the difference in the continuum absorption models of water vapor (see below). 

We also compare the planetary radiation and Bond albedos calculated with our radiative transfer model with those by previous studies to check the validity of our model (Fig. \ref{fig:Fpl_comp}). The radiation limit of our model is 280 $\mathrm{W/m^2}$, and is in close agreement with the value of 282 $\mathrm{W/m^2}$ obtained by \citet{Goldblatt13} with line-by-line calculations, with a difference of less than 1\%. As the surface temperature increases, the planetary radiation starts to rise for all the three models due to the temperature increase in the upper atmosphere. In this phase, our results agree well with those by \citet{Kopparapu13}, rather than by \citet{Goldblatt13}, probably because of the differences in the treatment of the continuum absorption of water vapor. \citet{PR11} indicated that the self continuum absorption of water vapor is underestimated in a MT\_CKD 2.4 model, which was used by  \citet{Goldblatt13}, especially in a wavenumber range from 2000 to 3000 $\mathrm{cm^{-1}}$, compared with the MT\_CKD 2.5 model used in our model and a BPS formalism used by \citet{Kopparapu13}. This likely explains why the planetary radiation starts to increase at the lower surface temperature in \citet{Goldblatt13}. The Bond albedos with our model are larger than those with the previous studies by about 0.05. This is probably because we underestimate the absorption and scattering processes at the wavenumbers larger than 30,000 $\mathrm{cm^{-1}}$.



\end{document}